\documentclass[a4paper,fleqn,usenatbib]{mnras}

\usepackage[T1]{fontenc}
\usepackage{ae,aecompl}

\usepackage{graphicx}	% Including figure files
\usepackage{amsmath}	% Advanced maths commands
\usepackage{amssymb}	% Extra maths symbols
\usepackage{mathrsfs}   % for \mathscr{...}

\usepackage[english]{babel}
\usepackage{xspace} % for \xspace

\usepackage{units}
\usepackage{wasysym} % for \astrosun

\usepackage{cosmopaperdefs}

\newcommand*{\cov}[1]{\operatorname{Cov}[{#1}]}
\newcommand*{\Cov}[1]{\operatorname{Cov}\left[{#1}\right]}
\newcommand*{\corr}[1]{\operatorname{Corr}[{#1}]}

\newcommand*{\Lp}[1][\ell]{\mathcal{L}_{#1}}
\newcommand*{\Lpbar}[1][\ell,\mu]{\mathcal{\bar L}_{#1}}
\newcommand*{\jl}[1][\ell]{j_{#1}}
\newcommand*{\jlbar}[1][\ell]{\bar \jmath_{#1}}

\newcommand*{\Nmock}{N_\mathrm{m}}
\newcommand*{\Nbins}{N_\mathrm{b}}
\newcommand*{\bgal}{b}
\newcommand*{\Skaiser}{S}
\newcommand*{\Minerva}{\textsc{Minerva}\xspace}

\title[Gaussian anisotropic clustering covariance]{Gaussian covariance matrices for anisotropic galaxy clustering measurements}

\author[J. N. Grieb et al.]{Jan Niklas Grieb,$^{1,2}$\thanks{E-mail: jgrieb@mpe.mpg.de (MPE)}
Ariel G. S\'anchez,$^{2}$
Salvador Salazar-Albornoz,$^{1,2}$ and\newauthor
Claudio Dalla Vecchia$^{3,4}$\\
$^{1}$Universit\"ats-Sternwarte M\"unchen, Ludwig-Maximilians-Universit\"at M\"unchen, Scheinerstra\ss{}e 1, 81679 M\"unchen, Germany\\
$^{2}$Max-Planck-Institut f\"ur extraterrestrische Physik, Postfach 1312, Giessenbachstr., 85741 Garching, Germany\\
$^{3}$Instituto de Astrof\'\i{}sica de Canarias, C/ V\'\i{}a L\'actea s/n, 38205 La Laguna, Tenerife, Spain\\
$^{4}$Departamento de Astrof\'\i{}sica, Universidad de La Laguna, Av{.} del Astrof\'\i{}sico Francisco S\'anchez s/n, 38206 La Laguna, Tenerife, Spain
}

\date{Accepted 2016 January 07. Received 2016 January 07; in original form 2015 September 14}

\pubyear{2016}

\begin{document}
\label{firstpage}
\pagerange{\pageref{firstpage}--\pageref{lastpage}}
\maketitle

\begin{abstract}
Measurements of the redshift-space galaxy clustering have been a prolific source of cosmological information in recent years.
Accurate covariance estimates are an essential step for the validation of galaxy clustering models of the redshift-space two-point statistics.
Usually, only a limited set of accurate N-body simulations is available.
Thus, assessing the data covariance is not possible or only leads to a noisy estimate.
Further, relying on simulated realisations of the survey data means that tests of the cosmology dependence of the covariance are expensive.
With these points in mind, this work presents a simple theoretical model for the linear covariance of anisotropic galaxy clustering observations with synthetic catalogues.
Considering the Legendre moments (`multipoles') of the two-point statistics and projections into wide bins of the line-of-sight parameter (`clustering wedges'), we describe the modelling of the covariance for these anisotropic clustering measurements for galaxy samples with a trivial geometry in the case of a Gaussian approximation of the clustering likelihood.
As main result of this paper, we give the explicit formulae for Fourier and configuration space covariance matrices.
To validate our model, we create synthetic HOD galaxy catalogues by populating the haloes of an ensemble of large-volume N-body simulations.
Using linear and non-linear input power spectra, we find very good agreement between the model predictions and the measurements on the synthetic catalogues in the quasi-linear regime.
\end{abstract}


%%%%%%%%%%%%%%%%%%%%%%%%%%%%%%%%%%%%%%%%%%%%%%%%%%

%%%%%%%%%%%%%%%%% BODY OF PAPER %%%%%%%%%%%%%%%%%%

%%%%% INTRODUCTION %%%%%

\section{Introduction}

Galaxy clustering observations are a wealthy source of cosmological information.
The three-dimensional distribution of galaxies on large scales is usually characterized by means of two-point statistics such
as the power spectrum (PS) or the two-point correlation function (2PCF).
Galaxy redshift surveys such as the 2dF Galaxy Redshift Survey \citep{Colless:2001gk}, Sloan Digital Sky Survey (SDSS) I and II \citep{York:2000gk}, and the Baryonic Oscillation Spectroscopic Survey \citep[BOSS, part of SDSS-III;][]{Dawson:2012va} have demonstrated the significance that these probes have for precision cosmology.
Spectroscopic surveys that are ongoing or starting soon are the extended Baryonic Oscillation Spectroscopic Survey \citep[eBOSS, part of SDSS-IV;][]{Dawson:2015wdb}
and the Hobby Eberly Telescope Dark Energy Experiment \citep{Hill:2008mv}.
In the future, next-generation experiments such as the Dark Energy Spectroscopic Instrument \citep{Levi:2013gra}, 4MOST \citep{deJong:2014}, the Subaru Prime Focus Spectrograph \citep{Ellis:2012rn} and the space-based \emph{Euclid} mission \citep{Laureijs:2011gra} will allow for an even more accurate exploration of galaxy clustering over a wide range of redshifts, giving access to invaluable cosmological information.

The strength of the clustering signal along the line-of-sight direction (LOS) differs from that in the transverse directions because of the peculiar
velocities of the galaxies.
This effect, dubbed redshift-space distortions \citep[RSD;][]{Kaiser:1987qv}, is mostly due to the large-scale infall 
of galaxies into gravitational potential wells and the non-linear velocity dispersion due to the orbital motion of galaxies in dense, collapsed structures.
By assuming a fiducial cosmology to convert redshifts to distances, measurements of the anisotropic clustering pattern caused by RSD provide information on the linear growth of structure  \citep{Sanchez:2013uxa,Beutler:2013yhm,Samushia:2013yga,Reid:2014iaa}.
Further, differences between the true cosmology of the Universe and the assumed fiducial one 
affect anisotropic clustering measurements through the so-called Alcock-Paczynski effect \citep{AP:1979}, which can be used in combination with
the signal from baryonic acoustic oscillations as a probe of the expansion history of the universe.
Modelling this effect allows for measurements of the angular diameter distance $D_\mathrm{A}(\bar z)$ to and the Hubble parameter
$H(\bar z)$ at the mean redshift $\bar z$ of the survey \citep[see e.g.][]{Anderson:2013zyy}.

A crucial ingredient of these cosmological analyses is an accurate model of the observed clustering statistics at cosmological, \ie, quasi-linear, scales, 
including the effect of non-linear evolution, galaxy bias and RSD.
The validation of such models requires precise estimates of the covariance matrix of the measured anisotropic power spectrum or two-point correlation function.
While the estimate of the clustering covariance matrix for the galaxy sample itself can be generated by \emph{brute-force} production of hundreds or thousands of synthetic realizations using fast approximate schemes for the formation of cosmological structures, model testing is better performed with the full non-linear covariance of the clustering statistic.
Hence, the model verification usually relies on large-volume N-body simulations of which only a few realizations can be produced due to limited run-time and memory.
Covariance estimates from the `mock' method suffer from the noise due to the finite number of realizations affecting the inverse of the covariance matrix, which propagates into the obtained constraints \citep{Taylor:2012kz,Dodelson:2013uaa,Percival:2013sga,Taylor:2014ota}.
The requirement to keep these error contributions small means that the number of mock realizations could reach the regime of a couple of thousands.
The complexity of the problem increases further if the dependence of the covariance on the cosmological model is to be analysed as well \citep[\eg, for the BAO covariance,][]{Labatie:2012ue}.
Then, the number of mock realisations significantly increases and the dependence of the covariance on any cosmological parameter can only be interpolated.

Alternative methods can be applied to reduce the required number of independent realizations.
\citet{Hamilton:2005dx} uses a reshuffling scheme for the phases of a periodic simulation to estimate the PS covariance from a single realization.
A similar resampling was used by \citet{Schneider:2011wf}.
Parametric models of the covariance matrix that only depend on a few free parameters would also be calibrated against a smaller number of simulations \citep{Pearson:2015gca}.
The noise in the obtained covariance matrix can also be reduced by means of a shrinkage estimation
\citep{Pope:2007vz} or covariance tapering \citep{Paz:2015kwa}.

In this work, we present the most simplistic theoretical framework for the modelling of anisotropic clustering covariance.
Previous studies showed that a Gaussian likelihood for the two-point statistics is sufficient for RSD experiments in the linear and quasi-linear regime \citep{Manera:2012sc}.
This paper aims at an extension of such covariance modelling for anisotropic clustering statistics in order to allow RSD analysis and model
verification using synthetic catalogues with periodic boundary conditions.
So far, the theoretical modelling of the covariance of anisotropic clustering statistics has not been studied in detail -- in contrast to the covariance of the monopole of 3D clustering statistics \citep{Feldman:1993ky,Smith:2007gi,Sanchez:2008iw} or the covariance of angular 2-point statistics \citep{Crocce:2010qi}.
In this paper we present the explicit formulae for the covariance of the multipoles as well as clustering wedges of the redshift-space galaxy PS and 2PCF for cubic mock catalogues.
For RSD and BAO studies, usually distance scales between $30 \; h^{-1} \, \unit{Mpc}$ and $180 \; h^{-1} \, \unit{Mpc}$ and wavenumbers between $0.02 \; h \, \unit{Mpc}^{-1}$ and $0.2 \; h \, \unit{Mpc}^{-1}$ are considered as relevant (quasi-linear regime of gravitational evolution, BAO features within this range).
Our model is verified on these scales using a set of large-volume N-body simulation whose haloes have been populated with galaxies mimicking the clustering statistics of real surveys by use of the halo occupation distribution \citep[HOD;][]{Peacock:2000qk,Scoccimarro:2000gm,Berlind:2001xk} technique.
The incorporation of the effects of a non-trivial survey geometry (described by the angular and radial selection of galaxies) is left for future work.

This paper is organized as follows:
the anisotropic galaxy clustering statistics are introduced in Section~\ref{sec:statistics}.
Measurements of their anisotropic moments and their covariance are presented in Section~\ref{sec:measurements}.
Section~\ref{sec:cov_model} presents our theoretical modelling of the anisotropic clustering covariance and gives the main results of this paper.
In Section~\ref{sec:model_validation} we validate our modelling using a set of large-scale N-body simulations which are described in detail in section~\ref{sec:minerva_simulations}.
We show that our model successfully gives a smooth representation of the data covariance of the simulations anisotropic clustering in section~\ref{sec:validation_with_simulations}.
Section~\ref{eq:conclusions} concludes our work.

%%%%% THEORY %%%%%

\section{The Methodology}
\label{sec:methodology}

This section contains a description of the galaxy two-point statistics and their measurements.
Considering the covariance of the two-point statistics, we describe how they are estimated from data and give a theoretical model for the Gaussian covariance matrix which is the main result of this paper.

\subsection{Galaxy two-point statistics}
\label{sec:statistics}

Galaxy clustering measurements are based on the galaxy overdensity field,
\begin{equation}
 \delta(\V x) \equiv \frac{n(\V x) - \bar n(\V x)}{\bar n (\V x)},
\end{equation}
where $n(\V x)$ denotes the actual number of galaxies around a point $\V x$, and $\bar n(\V x)$ is the expected galaxy number density.
Most clustering analyses focus on the power spectrum (PS) and its Fourier transform, the two-point correlation function (2PCF).
The 3D power spectrum $P(\V k)$ gives the covariance of the overdensity field in Fourier space,
\begin{equation}
 \label{eq:def_ps}
 \average{ \hat \delta(\V k) \, \hat \delta^\ast(\VPrime k)} = (2 \pi)^3 \delta_\mathrm{D}(\V k - \VPrime k) \left[ P(\V k) + \bar n^{-1} \right],
\end{equation}
where $\delta_\mathrm{D}(\V k - \VPrime k)$ is the 3D Dirac delta function and $\hat \delta^\ast(\V k)$ denotes the complex conjugate of the Fourier transform of the overdensity, $\hat \delta(\V k)$.
For this relation we took into account that the discrete galaxy positions of a given survey are sampled by a Poisson point process leading to a shot-noise contribution $\bar n^{-1}$ \citep{Feldman:1993ky}.
Redshift-space distortions depend on the line-of-sight (LOS) parameter $\mu$ which is defined as the cosine of the angle between the LOS and the separation vector of a galaxy pair.
None the less, $P(\V k)$ is still statistically invariant under rotations around the LOS.
Hence, we express the power spectrum in terms of the absolute wavenumber and the LOS parameter, $P(\V k) = P(k, \mu)$.

In order to have measurements with a significant signal-to-noise ratio, the anisotropic PS is usually projected onto multipole moments \citep{Padmanabhan:2008ag},
\begin{equation}
 \label{eq:ps_ell}
 P_\ell(k) \equiv \frac{2 \ell + 1}{2} \int_{-1}^1 \Lp(\mu) \, P(k, \mu) \dint \mu,
\end{equation}
where $\Lp(\mu)$ denotes the Legendre polynomial of order $\ell$.
On the other hand, clustering wedges, which have first been defined in configuration space by \citet{Kazin:2011xt}, are projections into wide bins of the LOS parameter,
\begin{equation}
 \label{eq:ps_w}
 P_{\mu_1}^{\mu_2}(k) \equiv \frac{1}{\mu_2 - \mu_1} \int_{\mu_1}^{\mu_2} P(k, \mu) \dint \mu.
\end{equation}
where $\mu_1$ and $\mu_2$ define the lower and upper limit, respectively, of non-intersecting wedges such that $\mu_2 - \mu_1 = \Delta \mu$.
The total number of wedges defines $\Delta \mu$.
We use the usual convention that in the case of only two wedges, these are labelled $P_\perp$ and $P_\parallel$ for the $\mu$-ranges $[0, 0.5]$ and $[0.5, 1]$, respectively.

Anisotropic clustering is often analysed in terms of the 2PCF defined as $\xi(\V s) \equiv \average{\delta(\V x) \, \delta(\V x + \V s)}$.
Hence, it is the configuration-space counterpart of the PS,
\begin{equation}
 \label{eq:xi_ell_from_ps}
 \xi(\V s) = \frac{1}{(2\pi)^3} \int P(\V k) \, \ft{\V k}{\V s} \, \dnx{3}{k}.
\end{equation}
The 2PCF can also be decomposed in terms of the Legendre multipoles \citep{Hamilton:1997kw}; the decomposition can be derived from the PS multipoles using
\begin{equation}
 \label{eq:xi_ell_from_ps_1d}
 \xi_\ell(s) = \frac{\ii^\ell}{2 \pi^2} \int_0^\infty P_\ell(k) \, \jl(ks) \, k^2 \dint k,
\end{equation}
where $\jl(x)$ is the spherical Bessel function of order $\ell$.
Clustering wedges in configuration space are defined completely analogous to their Fourier space counterparts,
\begin{equation}
 \label{eq:xi_w}
 \xi_{\mu_1}^{\mu_2}(s) \equiv \frac{1}{\mu_2 - \mu_1} \int_{\mu_1}^{\mu_2} \xi(s,\mu) \dint \mu.
\end{equation}
By convention, we use the labels $\xi_\perp$ and $\xi_\parallel$ for the case of two wedges.

The relation between 2PCF wedges and multipoles is
\begin{equation}
 \label{eq:xi_w_from_ell}
 \xi_{\mu_1}^{\mu_2}(s) = \sum_\ell \xi_\ell(s) \, \Lpbar,
\end{equation}
where $\Lpbar$ is the average of a Legendre polynomial over the $\mu$-range of the wedge,
\begin{equation}
 \label{eq:Lpbar}
 \Lpbar \equiv \frac{1}{\Delta \mu} \int_\mu^{\mu + \Delta \mu} \Lp(\mu) \dint \mu.
\end{equation}
This integral is evaluated in equation~(\ref{eq:Lpbar_eval}) in the appendix.

%%%%% PS and 2PCF Measurements %%%%%

\subsection[Measurements of the galaxy clustering covariance]{Measurements of the covariance matrix of two-point clustering measurements}
\label{sec:measurements}

We consider that the moments of the anisotropic PS have been measured to be $\hat P^i_x \equiv \hat P_x(k_i)$ in $\Nbins$ different wavenumber bins whose centres are at $k_i$.
The subscript $x$ can either refer to a multipole index $\ell$ or to a clustering wedge limited by some $\{ \mu_1, \mu_2 \}$.
Analogously, the 2PCF moments, $\hat \xi_x^i \equiv \hat \xi_x(r_i)$, have been measured in $\Nbins$ separation bins centred at $r_i$.

Assuming that $\hat \delta(\V k)$ is a Gaussian random field, the Fourier modes $\abs{\hat \delta(\V k)}^2$ follow a Rayleigh distribution \citep[see \eg][]{Kalus:2015lna}.
In the following, we assume that the number of Fourier modes observed is large enough to validate the assumption that the power spectrum follows a multi-variate Gaussian distribution with fixed covariance.\footnote{Deviations from the Gaussianity of the angular power spectrum likelihood have recently been analysed \citep{Sun:2013nna,Kalus:2015lna} and it was shown that constraints on primordial non-Gaussianity are affected by the negligence of the non-Gaussian contribution.
The logarithm of the spatial power spectrum was found to be closer to a Gaussian random variable at large scales than the power spectrum itself by \citet{Ross:2012sx}.
We neglect the dependence of the covariance on cosmological parameters as suggested by \citet{Carron:2012pw}.}

Given that the true inverse of covariance matrix, $\boldsymbol{\psi}$, is known and fixed, we can compute the likelihood of a model prediction for the clustering statistic, $P_x^i$ or $\xi_x^i$, by\footnote{For the 2PCF likelihood, replace the $P^i_x$ and $\hat P^i_x$ by $\xi^i_x$ and $\hat \xi^i_x$.}
\begin{equation}
 \label{eq:ps_likelihood}
 \mathscr{L} \left( P^i_x \mid \hat P^i_x, \boldsymbol{\psi} \right) = \frac{\abs{\boldsymbol{\psi}}}{\sqrt{2 \pi}} \exp \left[ - \frac 1 2 \chi^2(P^i_x, \hat P^i_x, \boldsymbol{\psi}) \right].
\end{equation}
Here, the \emph{log-likelihood} or $\chi^2$ function is given by
\begin{equation}
 \label{eq:chi_sqr}
 \chi^2(P^i_x, \hat P^i_x, \boldsymbol{\psi}) = \sum_{x,y} \sum_{i,j} \left( \hat P^i_x - P^i_x \right) \psi_{xyij} \left( \hat P^j_y - P^j_y \right),
\end{equation}
where $\psi_{xyij}$ are the elements of the precision matrix $\boldsymbol{\psi}$.

Usually the covariance matrix ${\mathbfss C}^P$ is estimated from a set of $\Nmock$ mock measurements, denoted ${}^{(n)}P^i_x$, where $n \in \{1, \ldots, \Nmock \}$, so that\footnote{The 2PCF covariance matrix $C^\xi_{xyij}$ is likewise estimated from mock measurements ${}^{(n)}\xi^i_x$.}
\begin{equation}
 C^P_{xyij} = \frac{1}{\Nmock - 1} \sum_n \left( {}^{(n)}P^i_x - \average{ P^i_x } \right) \left( {}^{(n)}P^j_y - \average{ P^j_y } \right).
\end{equation}
Here, the mean over all simulations is given by
\begin{equation}
 \average{ P^i_x } = \frac{1}{\Nmock} \sum_n {}^{(n)} P^i_x.
\end{equation}
In order to account for the noise in $({\mathbfss C}^P)^{-1}$, the likelihood of equation~(\ref{eq:ps_likelihood}) must be marginalized over the distribution of these uncertainties which, in the case in which $^{(n)} P^i_x$ are Gaussian random samples, follows an inverse Wishart distribution with the true precision matrix $\boldsymbol{\psi}$.
In most analyses however, $({\mathbfss C}^P)^{-1}$ is treated as the exact inverse covariance, resulting in known biases of the parameter estimates which must be corrected for \citep[see \eg][]{kaufman1967,Hartlap:2006kj,Percival:2013sga,Taylor:2014ota}.
The smaller the number of realizations, $\Nmock$, the larger these corrections are; in the extreme case of $\Nbins \ge \Nmock - 2$, the covariance matrix ${\mathbfss C}^P$ becomes singular.

We want to avoid generating and processing a large number of synthetic catalogues and hence aim for a smooth and non-singular estimate of the covariance matrix by theoretical modelling of the data covariance and its inverse.

%%%%% PS and 2PCF Covariance %%%%%

\subsection[Modelling of the galaxy clustering covariance]{Modelling of the covariance of galaxy two-point clustering measurements}
\label{sec:cov_model}

\subsubsection{Fourier Space: the Power Spectrum Covariance}

The assumption that the 2D power spectrum follows a Gaussian distribution leads to the following relation for the PS mode-by-mode covariance \citep[see][for the monopole]{Feldman:1993ky},
\begin{equation}
 \label{eq:ps_cov}
 \Cov{P(\V k), P(\VPrime k)} = \frac{2 \, (2 \pi)^3}{V_\mathrm{s}} \delta_\mathrm{D}(\V k - \VPrime k) \left[ P(k, \mu) + \bar n^{-1} \right]^2,
\end{equation}
where $V_\mathrm{s}$ is the volume of the sample.\footnote{In our case of a periodic box, the volume is given by $V_\mathrm{s} = L^3$, where $L$ is the side length of the box.
For a survey with a selection function given by a random catalogue with varying number density, the effective volume has to be estimated from the window function \citep{Bernstein:1993nb,dePutter:2011ah}.}
The Dirac delta function reflects the independence of Fourier modes for a random field with statistical translational invariance and in the absence of gravitational mode coupling.
The anisotropy of the PS is taken into account by the LOS dependency of $P(k, \mu)$.

Here, we neglected the trispectrum contribution \citep{Scoccimarro:1999kp} and the super-sample covariance (SSC; also called \emph{beat-coupling}).
Further, the fact that the local density estimate is obtained in the presence of power on scales larger than the survey size affects the power spectrum covariance of real surveys in the same manner as SSC, but the net effect only makes up 10\% of the original beat-coupling effect \citep{dePutter:2011ah}.
The contribution of the two latter effects to the PS covariance has been shown to become important at non-linear scales $k \gtrsim 0.1 \; h \, \unit{Mpc}^{-1}$ \citep{dePutter:2011ah,Takada:2013wfa,Li:2014sga}.
The mode coupling in the standard trispectrum term due to non-linear gravitational evolution is subdominant compared to SSC, but adds to it creating a plateau in the signal-to-noise ratio of the matter power spectrum in the mildly non-linear regime and beyond \citep{Carron:2014hja}.
The modelling of these effects is beyond the scope of this paper and their analysis is left for future work.

The theoretical covariance matrix for binned anisotropic PS measurements is obtained by averaging over the number of independent Fourier modes $\V k$ that contribute to each wavenumber bin.
We assume the bins to be centred at $k_i$ with width $\Delta k$.
For readability, the derivation of the following covariance expressions are presented in appendix~\ref{app:cov_proofs}.
A useful definition is the multipole expansion of the per-mode covariance given by
\begin{multline}
 \label{eq:ps_cov_ell_ell}
 \sigma^2_{\ell_1\ell_2}(k) \equiv \frac{(2 \ell_1 + 1) \, (2 \ell_2 + 1)}{V_\mathrm{s}} \\
 \times \int_{-1}^1 \left[ P(k, \mu) + \frac{1}{\bar n} \right]^2 \Lp[\ell_1](\mu) \, \Lp[\ell_2](\mu) \dint \mu.
\end{multline}
The normalization was chosen such that the pre-factor of $\sigma^2_{00}(k)$ is the usual $2/V_\mathrm{s}$ for an isotropic PS.
Equation~(\ref{eq:ps_cov_ell_ell_in_ps_ell}) in appendix~\ref{app:cov_proofs} provides the corresponding formula expressing the power spectrum in a multipole series.

As shown in section~\ref{app:ps_multipoles}, our ansatz for the bin-averaged PS multipole covariance $C^P_{{\ell_1}{\ell_2}}(k_i,k_j) \equiv \Cov{P^i_{\ell_1}, P^j_{l_2}}$ yields
\begin{equation}
 \label{eq:ps_ell_cov_bin}
 C^P_{{\ell_1}{\ell_2}}(k_i,k_j) = \frac{2 \, (2 \pi)^4}{V_{k_i}^2} \delta_{ij} \int_{k_i-\Delta k/2}^{k_i+\Delta k/2} \sigma^2_{\ell_1\ell_2}(k) \, k^2 \dint k,
\end{equation}
where the volume of the shell in $k$-space is $V_{k_i} = 4 \pi [(k_i + \Delta k/2)^3 - (k_i - \Delta k/2)^3] / 3$.
The full multipole expansion presented in the appendix shows that the multipole covariance matrix has terms which mix all multipole contributions.
This implies that the monopole covariance is not only given by terms depending on $(P_0(k) + \bar n^{-1})^2$, but also contains contributions from all higher-order multipoles as well.

Second, we consider clustering wedges $P_{\mu_1}^{\mu_2}(k)$ as defined by equation~(\ref{eq:ps_w}).
We denote the measurement of $P_{\mu}^{\mu+\Delta \mu}$ in the wavenumber bin centred around $k_i$ as $P_\mu^i$.
Then, the theoretical covariance of the bin-averaged wedges, $C^P_{\mu\mu'}(k_i, k_j) \equiv \Cov{P^i_{\mu}, P^j_{\mu'}}$, is given by (see section~\ref{app:ps_wedges})
\begin{multline}
 \label{eq:ps_w_cov_bin}
 C^P_{\mu\mu'}(k_i, k_j) = \frac{4 \, (2 \pi)^4}{V_{k_i}^2 \, (\Delta \mu)^2} \delta_{ij} \, \delta_{\mu \mu'} \\
 \times \int_{k_i-\Delta k/2}^{k_i+\Delta k/2} \int_\mu^{\mu+\Delta \mu} \left[ P(k, \tilde \mu) + \frac{1}{\bar n} \right]^2 \dint \tilde \mu \, k^2 \dint k,
\end{multline}
where $\delta_{\mu \mu'}$ reflects that the wedges are discrete and non-intersecting.
In contrast to the multipoles, different Fourier space wedges are not correlated in our linear Gaussian theory, even after integration over the wavenumber bin.

\subsubsection{Configuration Space: the 2PCF Covariance}

As in the previous section, we only present here the final results and give the full derivations in section~\ref{app:2pcf_multipoles}.
By defining the per-mode multipole covariance as in equation~(\ref{eq:ps_cov_ell_ell}), we can write the bin-averaged 2PCF multipole covariance matrix, $C^\xi_{{\ell_1}{\ell_2}}(s_i,s_j) \equiv \Cov{\xi^i_{\ell_1}, \xi^j_{l_2}}$, as
\begin{equation}
 \label{eq:2pcf_ell_cov_bin}
 C^\xi_{{\ell_1}{\ell_2}}(s_i,s_j) = \frac{\ii^{\ell_1+\ell_2}}{2 \pi^2} \int_0^\infty k^2 \, \sigma^2_{\ell_1\ell_2}(k) \, \jlbar[\ell_1](ks_i) \, \jlbar[\ell_2](ks_j) \dint k,
\end{equation}
where $\jlbar(ks_i)$ is the bin-averaged spherical Bessel functions as defined in equation~(\ref{eq:jlbar}).
The bin average has been shown to be necessary in order not to overestimate the 2PCF covariance \citep{Cohn:2005ex,Sanchez:2008iw}.

In section~\ref{app:2pcf_wedges}, we derive the covariance of the 2PCF clustering wedges, $C^\xi_{\mu\mu'}(s_i, s_j) \equiv \Cov{\xi^i_{\mu}, \xi^j_{\mu'}}$, to be
\begin{multline}
 \label{eq:2pcf_w_cov_bin}
 C^\xi_{\mu\mu'}(s_i, s_j) = \sum_{\ell_1, \ell_2} \frac{\ii^{\ell_1+\ell_2}}{2 \pi^2} \Lpbar[\ell_1,\mu] \, \Lpbar[\ell_2,\mu'] \\
 \times \int_0^\infty k^2 \, \sigma^2_{\ell_1\ell_2}(k) \, \jlbar[\ell_1](ks_i) \, \jlbar[\ell_2](ks_j) \dint k.
\end{multline}
Here, $\Lpbar$ is the average of a Legendre polynomial over the $\mu$-range of the wedge as defined in equation~(\ref{eq:Lpbar}).

%%%%% VALIDATION %%%%%

\section{Validation of the Model}
\label{sec:model_validation}

In this section, we first describe the set of N-body simulations that we will use to validate our model of the covariance of anisotropic galaxy clustering two-point statistics.
Next, we show and discuss the agreement between the theoretical prediction and the results from the simulations.
We specifically address the question of whether the predictions presented in the previous section are precise enough to allow performance test of RSD models for the clustering statistics of current galaxy surveys.

%%%%% Minerva Simulations %%%%%

\subsection{The N-body Simulations and Galaxy Catalogues}
\label{sec:minerva_simulations}

\begin{table}
 \centering
 \caption{The cosmological parameters of our set of \Minerva simulations.}
 \label{tab:cosmo_params}
  \begin{tabular}{llllll}
   \hline
   parameter & $\Omega_\mathrm{m}$ & $\Omega_\Lambda$ & $h$   & $n_\mathrm{s}$ & $\sigma_8$ \\
   value     & 0.285               & 0.715            & 0.695 & 0.9632         & 0.828 \\
   \hline
  \end{tabular}
\end{table}

\begin{figure}
 \includegraphics[width=\columnwidth]{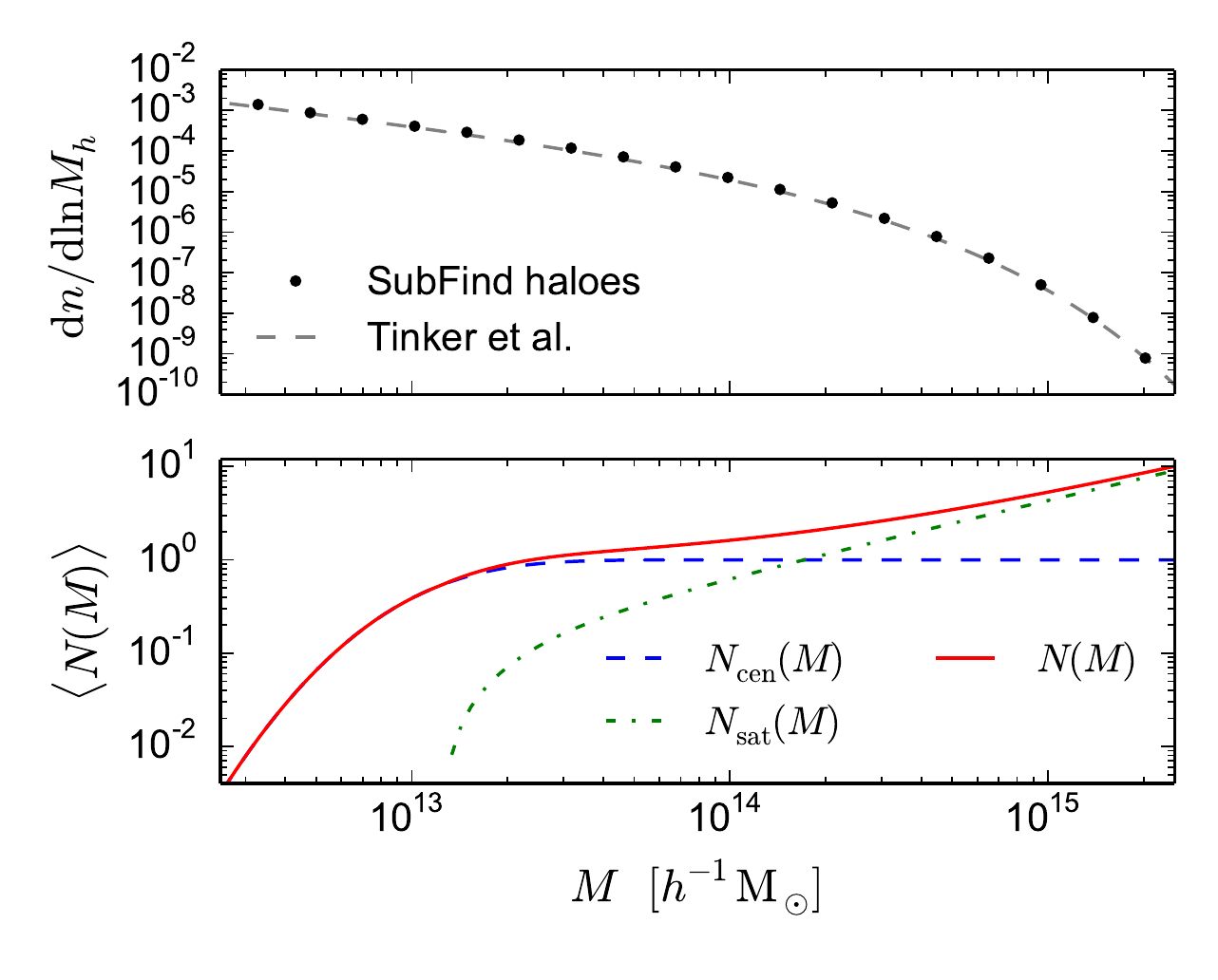}
 \caption{\emph{Upper panel:} the mean halo mass function of the SubFind output for our \Minerva snapshots at $z = 0.57$ (black points) and the prediction (gray dashed line) based on the recipe of \citet{Tinker:2008ff}.
  \emph{Lower panel:} the mean galaxy occupation function as defined in equation~(\ref{eq:HOD_tot}), $\average{N(M)}$ (red solid line), and its decomposition into central (blue dashed line) and satellite components (green dotted line) for the parameters given in Table~\ref{tab:HOD_params}.}
 \label{fig:minerva_HOD}
\end{figure}

The \Minerva simulations are a set of 100 N-body simulations run using \textsc{Gadget}\footnote{The latest public release is \textsc{Gadget-2} which is available at \url{http://www.gadgetcode.org/}.} \citep[last described in][]{Springel:2005mi} with $1000^3$ dark matter (DM) particles per realization in a cubic box of side length $1500 \; h^{-1} \, \unit{Mpc}$ with periodic boundary conditions.
The simulations were started at redshift $z_\mathrm{ini} = 63$ using 2LPT initial conditions.
The input linear power spectrum for the initial conditions was calculated using \textsc{Camb} \citep{Lewis:1999bs} for the cosmological parameters chosen for our set of simulations (see Table~\ref{tab:cosmo_params}),
which match the best-fitting $\Lambda$CDM model of the WMAP9 + BOSS DR9 $\xi(r)$ analysis \citep[Table I]{Sanchez:2013uxa}.
The positions and velocities of the DM particles were stored for five output redshifts $z \in \{2.0, 1.0, 0.57, 0.3, 0\}$, jointly with the halo positions and velocities found by a friends-of-friend halo finder applied with the standard linking length equal to 0.2 of the mean inter-particle separation.
In a post-processing step, \textsc{SubFind} \cite[section~4.2]{Springel:2000qu} was run to generate the final halo catalogues for each realization and output redshift.
The mean halo mass function for $z = 0.57$ is plotted in the upper part of Fig{.}~\ref{fig:minerva_HOD}.
The minimum resolved halo mass is $m_\mathrm{min} = 2.67 \cdot 10^{12} \; h^{-1} \, \unit{M}_\Sun$.

\begin{table}
 \centering
 \caption{The parameters of the HOD model of equations~(\ref{eq:HOD_tot}) to (\ref{eq:HOD_sat}) defining our `CMASS-like' galaxy sample at $z=0.57$.
  All masses are in units of $h^{-1} \, \unit{M}_\Sun$.}
 \label{tab:HOD_params}
 \begin{tabular}{lr}
  \hline
  parameter & value \\
  \hline
  $\log_{10}(M_\mathrm{min})$ & 13.07  \\
  $\log_{10}(M_0)$            & 13.1   \\
  $\log_{10}(M_1')$           & 14.2   \\
  $\sigma_{\log_{10} M}$      &  0.347 \\
  $\alpha$                    &  0.8   \\
  \hline
 \end{tabular}
\end{table}

The volume of each realization is large enough to allow for the analysis of anisotropic galaxy clustering probes with a precision comparable to present-day galaxy redshift surveys.
In order to generate a galaxy sample comparable to the CMASS sample of BOSS, we populate the haloes and subhaloes of each simulation with galaxies according to a halo occupation distribution (HOD) model with suitable parameters.
In such HOD models the average number $N$ of synthetic galaxies in haloes of mass $M$ is given by the mean occupation function $N(M)$.
We follow the parametrization of \citet{Zheng:2007zg} and decompose this function into contributions from central and satellite galaxies,
\begin{equation}
 \label{eq:HOD_tot}
 \average{N(M)} = \average{N_\mathrm{cen}(M)} + \average{N_\mathrm{sat}(M)},
\end{equation}
where the mean central occupation function,
\begin{equation}
 \label{eq:HOD_cen}
 \average{N_\mathrm{cen}(M)} = \frac 1 2 \left[ 1 + \erf \left( \frac{\log M - \log M_\mathrm{min}}{\sigma_{\log M}} \right) \right],
\end{equation}
has a smooth cut-off at $M_\mathrm{min}$ modelled by an error function with relative scale $\sigma_{\log M}$ to describe the scatter between the galaxy luminosity and mass.
By setting the satellites contribution to
\begin{equation}
 \label{eq:HOD_sat}
 \average{N_\mathrm{sat}(M)} =  \average{N_\mathrm{cen}(M)} \left( \frac{M - M_0}{M_1'} \right)^\alpha,
\end{equation}
we only assign satellites to haloes that are already populated by a central galaxy.
The satellite galaxies are sampled with a Poisson distribution with the mean set by equation~(\ref{eq:HOD_sat}) where the cut-off mass scale is $M_0$, the normalization mass scale is $M_1'$, and the power-law slope is $\alpha$.
If a central galaxy is assigned to a halo, its position and velocity are derived from the most-bounded DM particle in that halo.
The positions and velocities of satellite galaxies are drawn from random DM particles associated to the hosting halo.

The parameters chosen for our HOD sample at redshift $z = 0.57$, the mean redshift of the BOSS CMASS sample, are given in Table~\ref{tab:HOD_params};
the resulting mean galaxy occupation function is shown in the lower panel of Fig{.}~\ref{fig:minerva_HOD}.
The final synthetic galaxy catalogues have a mean galaxy density of $\bar n \approx 4 \sciexp{-4} \; h^3 \, \unit{Mpc}^{-3}$.
By comparing the redshift-space 2PCF monopole of the HOD sample with the real-space one of the DM particles, we find a linear bias of $b^2 = 4.02$ derived from the ratio at pair separations $40 \; h^{-1} \, \unit{Mpc} \le s \le 60 \; h^{-1} \, \unit{Mpc}$.
Using the ratio of the real- and redshift-space monopole of the HOD sample at these separations, we find that the Kaiser factor is $\Skaiser = 1.28$, in perfect agreement with the theoretical value of $S = 1 + \frac 2 3 \beta + \frac 1 5 \beta^2$ \citep{Kaiser:1987qv} derived from the growth factor $f = 0.76$ for the \Minerva cosmology.
Here, we used $\beta = f / \bgal$, where $f$ is the growth rate, i{.}e{.}, the logarithmic derivative of the growth function w{.}r{.}t{.} the scale factor.

\begin{figure}
 \includegraphics[width=\columnwidth]{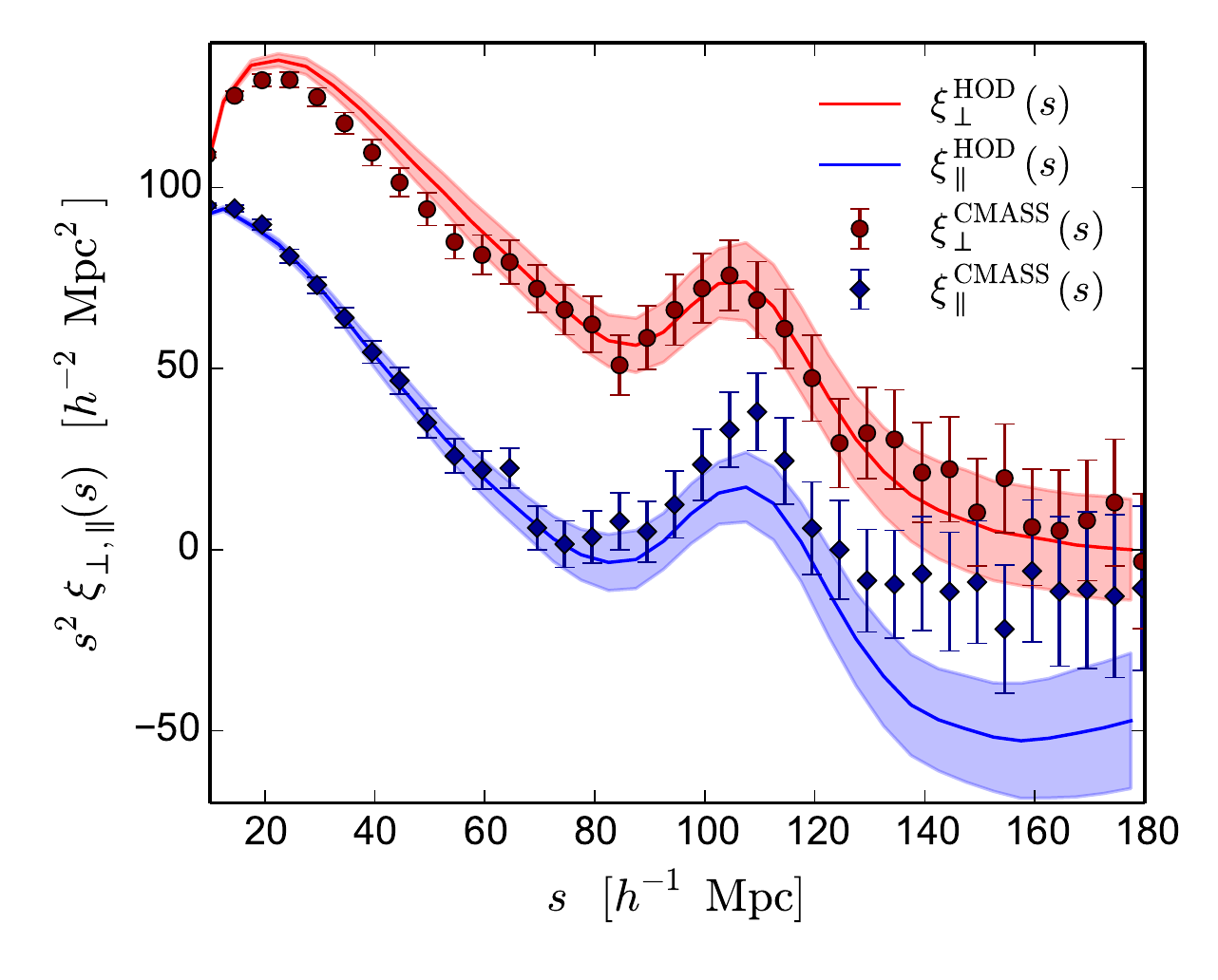}
 \caption{Comparison of the 2PCF clustering wedges $\xi^\mathrm{HOD}_\perp(s)$ and $\xi^\mathrm{HOD}_\parallel(s)$ of our HOD sample (red and blue solid lines, respectively, standard deviation over our 100 realizations indicated by the filled region) compared with the corresponding measurements from the CMASS sample of BOSS DR11 (transverse wedge: circles, parallel wedge: diamonds) by \citet{Sanchez:2013tga}.}
 \label{fig:minerva_2PCF_2w}
\end{figure}

The agreement between the anisotropic clustering of the HOD galaxy sample and the CMASS sample of BOSS DR11 can be seen in
Fig{.}~\ref{fig:minerva_2PCF_2w}, where we show the comparison of the 2PCF clustering wedges of both samples.

No super-survey modes have been taken into account for our simulations \citep[e{.}g{.} by use of the separate universe response,][]{Li:2014sga} in agreement with the absence of those modes in our modelling of the PS covariance in equation~(\ref{eq:ps_cov}).

\subsection{The two-dimensional power spectrum}

\begin{figure}
 \includegraphics[width=\columnwidth]{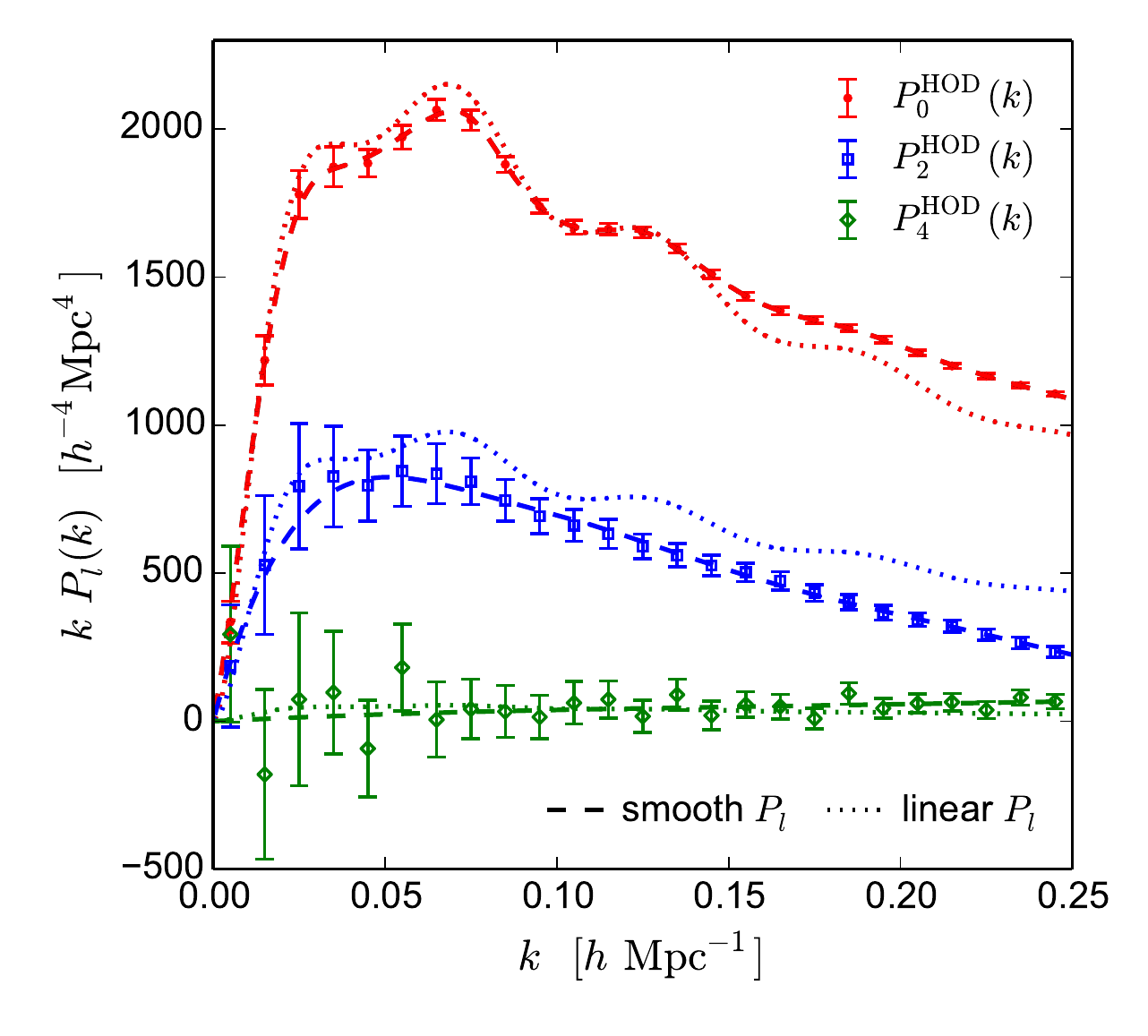}
 \caption{Comparison of the mean power spectrum multipoles of our HOD sample (points) with the linear-theory predictions, $P^\mathrm{lin}_l(k)$, of 
  equation~(\ref{eq:lin_Pk}) (dotted lines) and the smoothed interpolation, $P^\mathrm{smooth}_l(k)$ (dashed lines).
  Both the linear and smoothed multipoles are taken as input for linear and non-linear predictions of the Gaussian covariance.}
 \label{fig:minerva_PS_l_smoothed}
\end{figure}

As the simplest case, we assume a linear prediction for the 2D galaxy power spectrum where redshift-space anisotropies are caused by the linear Kaiser effect \citep{Kaiser:1987qv},
\begin{equation}
 \label{eq:lin_Pk}
 P^\mathrm{lin}(k,\mu) = \bgal^2 P_\mathrm{L}(k) \, (1 + \beta^2 \mu^2)^2,
\end{equation}
where $P_\mathrm{L}(k)$ is the linear matter power spectrum, and $\bgal$ is the linear galaxy bias.

With these assumptions, the only non-vanishing PS multipoles are the monopole, quadrupole, and hexadecapole.

Fig~\ref{fig:minerva_PS_l_smoothed} shows the mean PS monopole, quadrupole and hexadecapole from the \Minerva simulations.
The linear theory definition of equation~(\ref{eq:lin_Pk}), shown by the dotted lines, gives an inadequate description of the anisotropic
galaxy power spectrum in the quasi-linear regime.
To improve upon this description we perform a smoothing spline interpolation of our mean PS multipole measurements, $P^\mathrm{smooth}_l(k)$, shown by the dashed lines in Fig~\ref{fig:minerva_PS_l_smoothed}, to create a noiseless non-linear power spectrum that we can use for covariance predictions,
\begin{equation}
 \label{eq:smooth_Pk}
 P^\mathrm{nl}(k,\mu) = P^\mathrm{smooth}_0(k) + P^\mathrm{smooth}_2(k) \, \Lp[2](\mu) + P^\mathrm{smooth}_4(k) \, \Lp[4](\mu).
\end{equation}
For the estimation of the smoothing length, we take the measured dispersion of the PS multipoles into account.
The BAO wiggles in the smoothed quadrupole have been slightly damped by this procedure but, due to the small signal-to-noise ratio of the BAO feature in the quadrupole, this does not affect the predicted covariance.

In the same way as the linear model of equation~(\ref{eq:lin_Pk}), by definition this ansatz also has only monopole, quadrupole and 
hexadecapole contributions. Higher-order multipoles are negligible for $k \lesssim 0.25 \; h \, \unit{Mpc}^{-1}$.
In the following, we will use these two power spectrum models for the analysis and validation of our Gaussian covariance predictions (referred to as `lin' and `smooth', respectively).
In the case of an application of the Gaussian covariance model for the use of RSD model performance tests, the smoothed power spectrum can be replaced by a preliminary fit of the RSD model to the data.

%%%%% Covariance Validation %%%%%

\subsection[Accuracy of the Gaussian Covariance Model]{Validation of the Accuracy of the Gaussian Covariance Model}
\label{sec:validation_with_simulations}

\begin{table}
 \centering
 \caption{The binning choices for the validation of our covariance model.}
 \label{tab:bin_conf}
 \begin{tabular}{lccccc}
  \hline
  space & range        & $\Delta$ small & unit           & number of bins \\
        &              & $\Delta$ large &                &    \\
  \hline
  $s$   & $0$ - $180$  & $5$            & $h^{-1} \, \unit{Mpc}$ & 36 \\
        &              & $15$           &                        & 12 \\
  $k$   & $0$ - $0.25$ & $0.005$        & $h \, \unit{Mpc}^{-1}$ & 50 \\
        &              & $0.010$        &                        & 25 \\
  \hline
 \end{tabular}
\end{table}

In order to compare the theoretical predictions presented in section~\ref{sec:cov_model} with the noisy estimates from the \Minerva simulations, we chose two different binning configurations: one leading to an invertible covariance matrix (in case of multipoles up to the hexadecapole or three clustering wedges) denoted `large' and one appropriate for fitting of CMASS-like measurements (for which the data covariance matrix obtained from 100 catalogues is singular) denoted `small'.
These setups are listed in Table~\ref{tab:bin_conf}.

The measurements of the redshift-space two-point clustering assume a plane-parallel LOS along one of the axis of the simulation box (`distant-observer approximation').
In order to reduce the level of noise in the mean and covariance of the measurements, we measure the two-point statistic of every realization by assuming the LOS to be parallel to each of the three different axes and then average the results.

Due to this averaging over the three LOS axes, the error on the covariance cannot easily be predicted analytically.
Thus, we measure the error on the covariance matrix ${\mathbfss C}^P$ as jackknife estimate from the same set of mock measurements,
\begin{equation}
 \label{eq:jackknife}
 \left( \Delta C^P_{xyij} \right)^2 = \frac{\Nmock - 1}{\Nmock} \sum_m \left( ^{(m)}C^P_{xyij} - C^P_{xyij} \right)^2,
\end{equation}
where $^{(m)}C^P_{xyij}$ is the covariance estimate based on leaving out the $m$th realization,
\begin{equation}
 ^{(m)}C^P_{xyij} = \frac{1}{\Nmock - 1} \sum_{n \neq m} \left( {}^{(n)}P^i_x - \average{ P^i_x } \right) \left( {}^{(n)}P^j_y - \average{ P^j_y } \right).
\end{equation}
The error on the  covariance matrix of all configuration space measurements, $C^\xi_{xyij}$, are obtained in an analogous way.
We checked that the jackknife estimate of the error on the covariance matrix for the case of a single LOS axis is in close agreement with the theoretical prediction given in \citet{Taylor:2012kz}.

The calculation of the likelihood in equation~(\ref{eq:ps_likelihood}) requires the estimation of the inverse of the covariance matrix, the precision matrix.
If this matrix is taken to be the inverse of the data covariance, it is biased due to the sample variance.
The level of noise is quite high in our case due to the small number of realizations and large number of measurements bins.
Since for many analysis configurations, the data covariance is even singular, we do not comment on the accuracy of the precision matrix here.
We shortly discuss the inverse data covariance matrix from our set of simulations for one particular binning scheme in appendix~\ref{app:prec_mat}.

\subsubsection{Fourier Space Covariance}

\begin{figure}
 \includegraphics[width=\columnwidth]{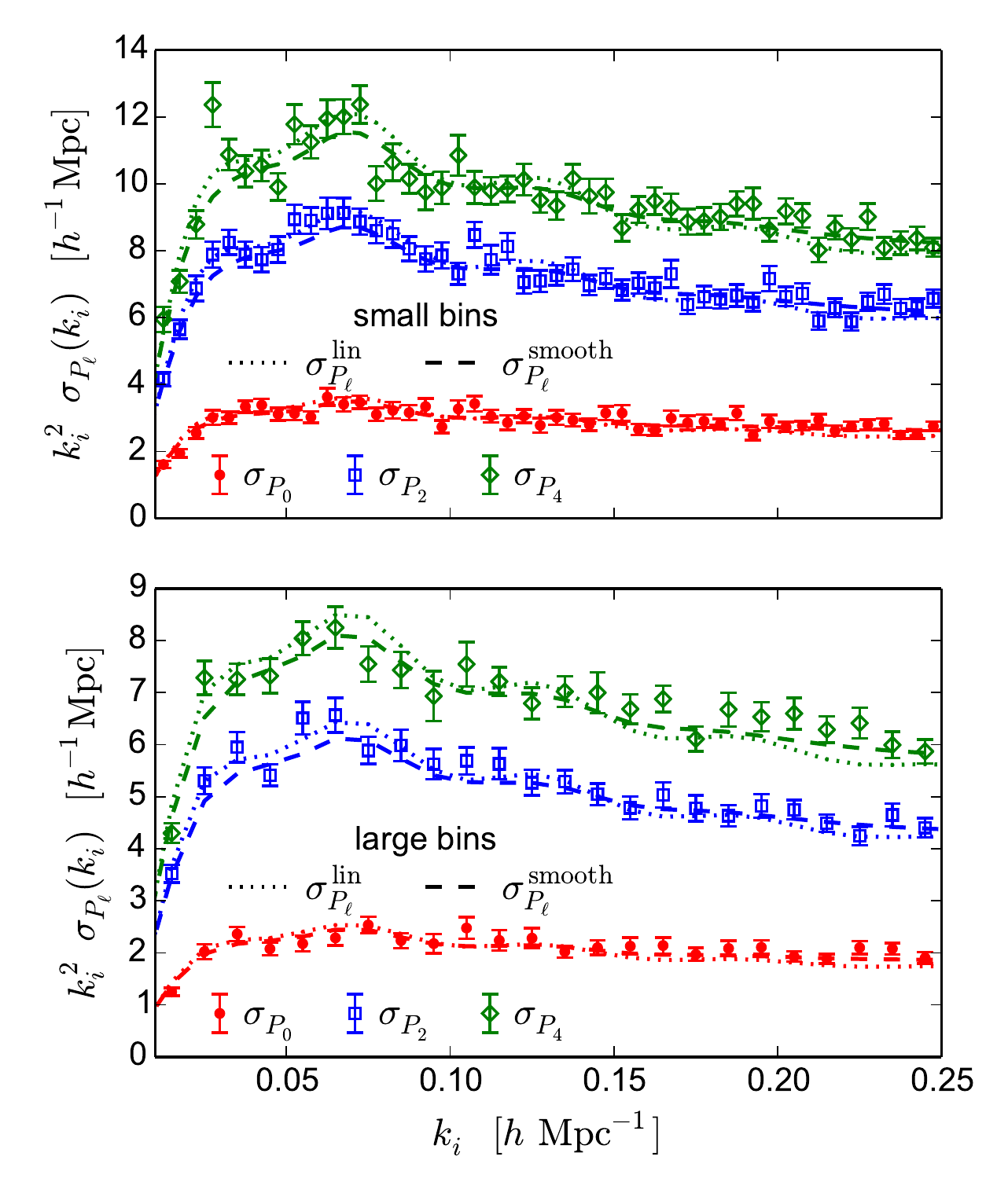}
 \caption{Comparison of the dispersion of the PS multipoles, $\sigma_{P_\ell}(k_i)$, of our HOD realizations (points) with the Gaussian predictions from equation~(\ref{eq:ps_ell_cov_bin}) for the small (upper panel) and large (lower panel) binning schemes.
  The dotted lines were derived using the linear model of equation~(\ref{eq:lin_Pk}), while the dashed lines correspond to the smoothed non-linear recipe of equation~(\ref{eq:smooth_Pk}).
}
 \label{fig:minerva_PS_ell_cov}
\end{figure}

The predicted covariance of the PS multipoles is compared against the data covariance measured from our 100 simulations in Fig{.}~\ref{fig:minerva_PS_ell_cov} for the small (upper panel) and large (lower panel) binning setup.
We show the dispersion of PS multipoles given by $\sigma^P_\ell(k_i) \equiv [C^P_{\ell\ell}(k_i,k_i)]^{1/2}$.
The error bar is likewise given by $\Delta \sigma^P_\ell(k_i) \equiv [\Delta C^P_{\ell\ell}(k_i,k_i)]^{1/2}$, where $\Delta C^P_{\ell\ell}(k_i,k_i)$ is estimated with the jackknife estimator given in equation~(\ref{eq:jackknife}).
The difference between linear and non-linear predictions is small for the monopole and quadrupole dispersion.
However, the linear prediction of the hexadecapole dispersion slightly overpredicts (underpredicts) the actual dispersion in the data for lower (higher) $k$, respectively, which is better matched by the results obtained from the smoothed multipole measurements.
In appendix~\ref{app:ps_multipoles}, we comment on the increase of the higher-order multipole variance showing predictions $\sigma_{P_6}$ and $\sigma_{P_8}$.

\begin{figure}
 \includegraphics[width=\columnwidth,page=2]{images/minerva_ps_multipoles_cov_mat.pdf}
 \includegraphics[width=\columnwidth,page=1]{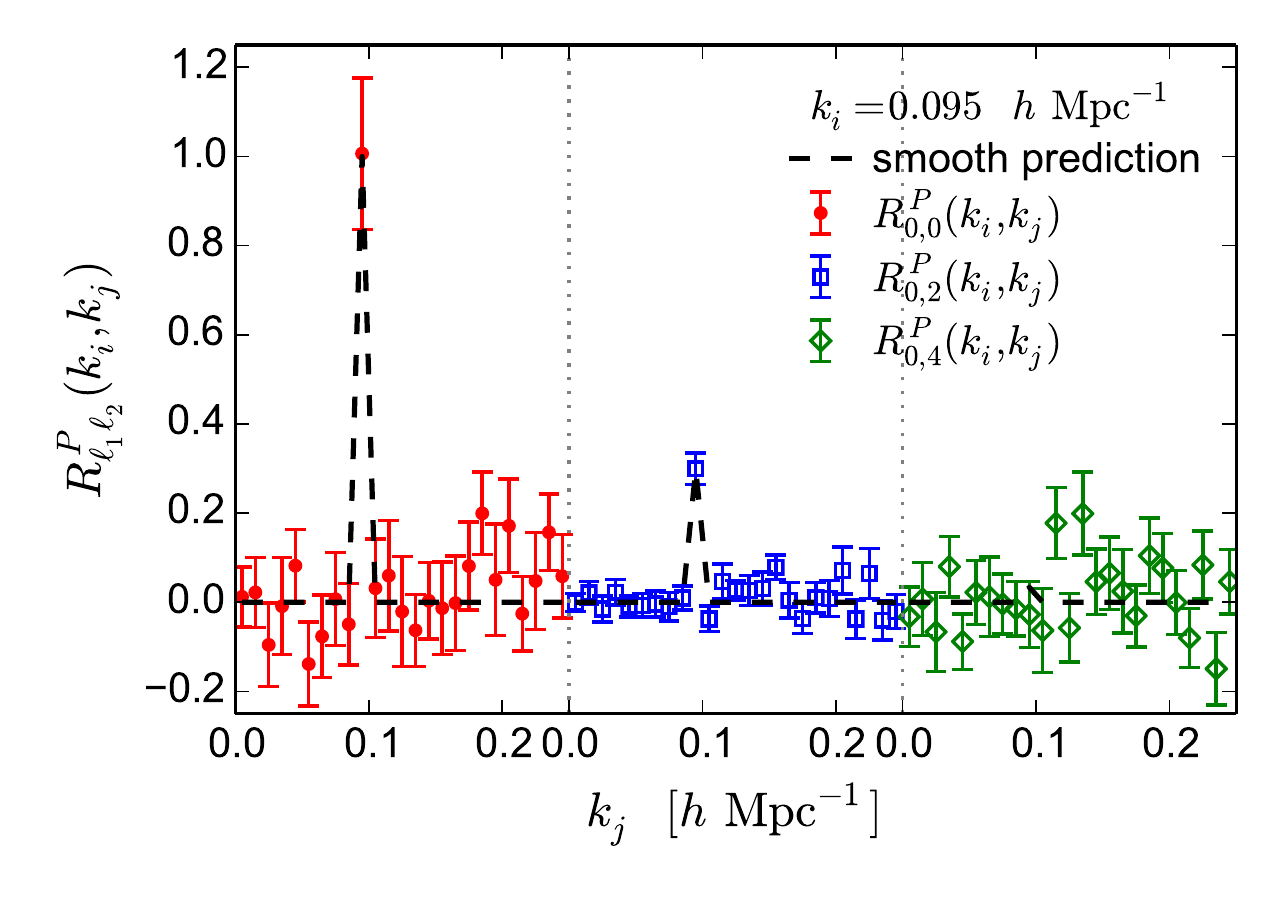}
 \caption{\emph{Upper panel:} The full correlation matrix of the PS multipoles, $R^P_{\ell_1\ell_2}(k_i,k_j) = C^P_{\ell_1\ell_2}(k_i,k_j) [\sigma_{P^\mathrm{nl}_{\ell_1}}(k_i) \, \sigma_{P^\mathrm{nl}_{\ell_2}}(k_j)]^{-1}$ (normalized by the theoretical prediction), shows a dominant diagonal and a significant cross-correlation between monopole and quadrupole (shown here: large binning).
  \emph{Lower panel:} Cut through the correlation matrix for $\ell_1 = 0$ at $k_i = 0.095 \; h \,\unit{Mpc}^{-1}$.
  The correlation contamination from physical effects not accounted for by our model and from noise is at the $15\%$ level and well within the error bars.}
 \label{fig:minerva_PS_ell_cov_mat}
\end{figure}

The measurements of the off-diagonal terms $C^P_{\ell_1,\ell_2}(k_i, k_j)$ (for $\ell_1 \neq \ell_2$ and/or $k_i \neq k_j$) suffer from low signal-to-noise and hence show larger relative scatter than the dispersion terms.
As the differences in our linear and non-linear modelling are much smaller than the variations in the measurements we find equally good agreement between the two predictions and our simulation data.
Because of the scatter, we cannot identify additional contributions such as non-gaussianities in our covariance measurements.
To highlight the full covariance properties, we define the correlation matrix for PS multipoles as
\begin{equation}
  R^P_{\ell_1\ell_2}(k_i,k_j) = C^P_{\ell_1\ell_2}(k_i,k_j) \, \left[ \sigma_{P^\mathrm{nl}_{\ell_1}}(k_i) \, \sigma_{P^\mathrm{nl}_{\ell_2}}(k_j) \right]^{-1}.
\end{equation}
The upper panel of Fig{.}~\ref{fig:minerva_PS_ell_cov_mat} we show a color representation of the data correlation matrix (upper triangular part) next to the theoretical prediction (lower triangular part).
The lower panel shows a cut through the matrix.
Note that we normalized the data covariance by the theoretical dispersion obtained from the smooth non-linear power spectrum,
$\sigma_{P^{\mathrm nl}_{\ell}}$, so that there is no further noise contamination.
The level of unaccounted data correlation in the off-diagonal terms is up to 20\%, but this is most likely due to noise contamination instead of systematics not included in our Gaussian ansatz.
The noise level for the monopole is larger than for the higher-order multipoles because of the average over the LOS directions which reduces the noise of the RSD but not of the real-space matter clustering.

\begin{figure}
 \includegraphics[width=\columnwidth]{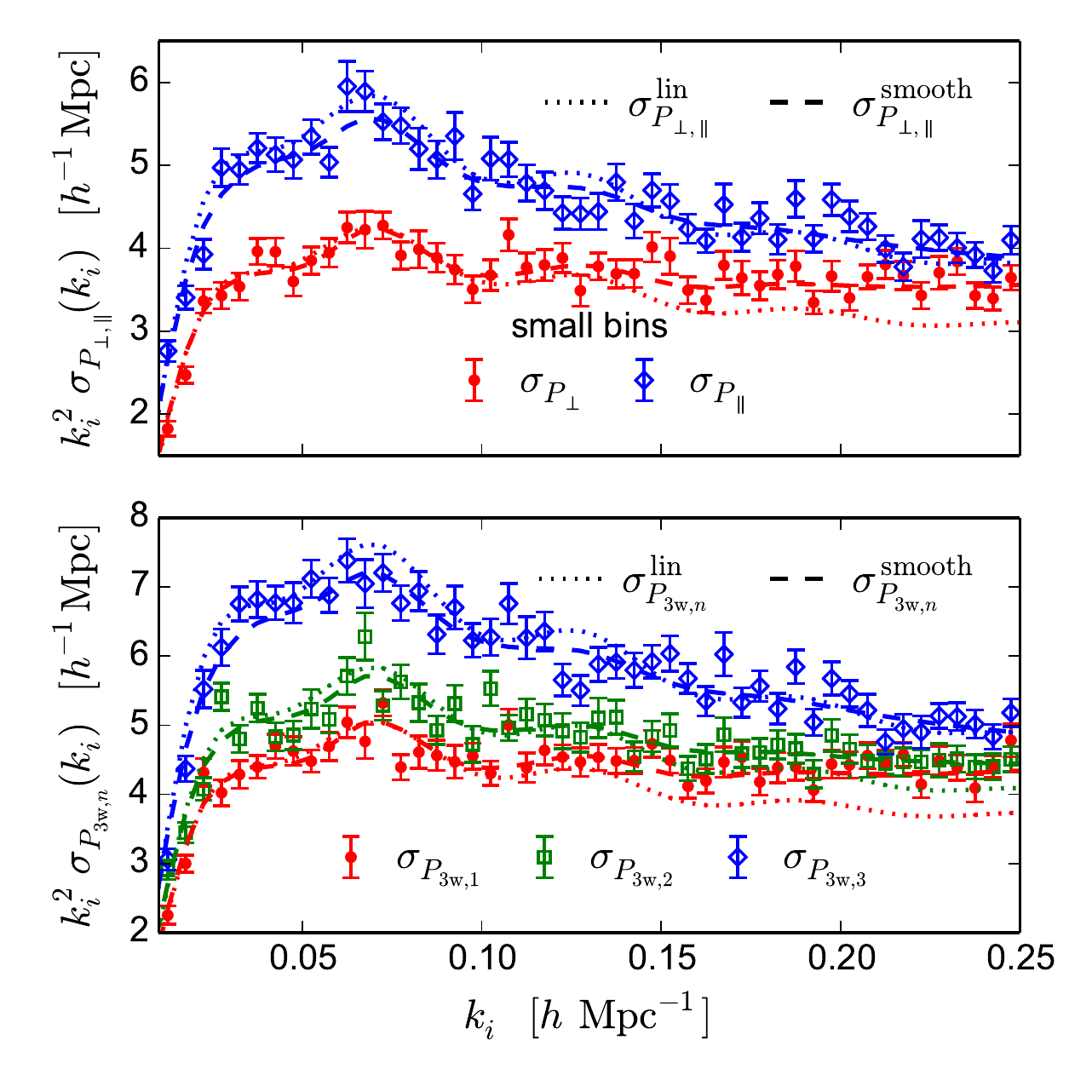}
 \caption{Comparison of the dispersion of the PS wedges of our HOD realizations (points) for the small binning configuration with the Gaussian predictions from equation~(\ref{eq:ps_w_cov_bin}) for the linear (dotted lines) and smoothed non-linear (dashed lines) input power spectra.
 The upper and lower panels show the cases of two wedges, $\sigma_{P_{\perp,\parallel}}(k_i)$, and three wedges, $\sigma_{P_{3\mathrm{w},n}}(k_i)$, respectively.}
 \label{fig:minerva_PS_w_cov_small}
\end{figure}

\begin{figure}
 \includegraphics[width=\columnwidth]{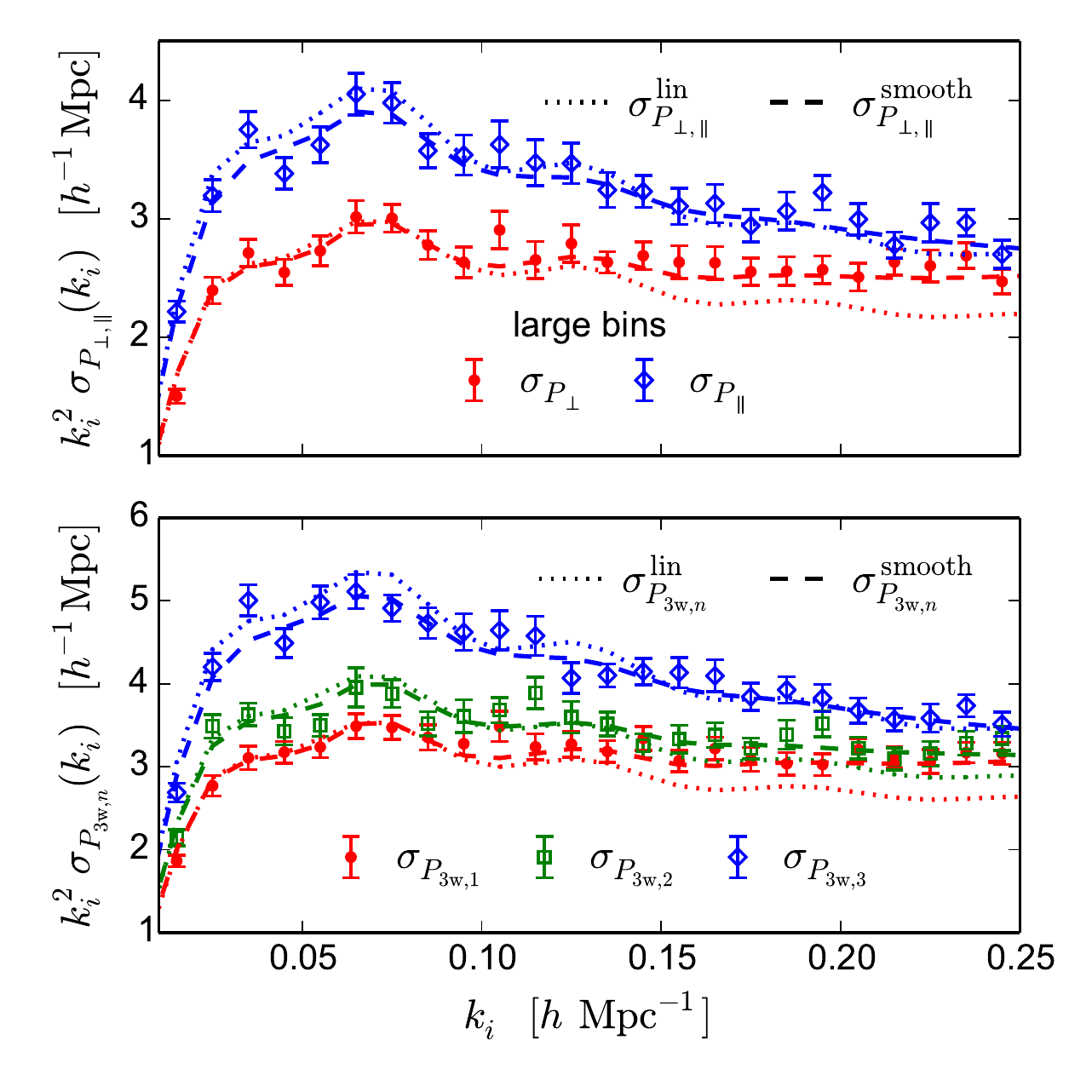}
 \caption{Comparison of the rms of the PS wedges of our HOD realizations (points) for the large binning configuration with the Gaussian predictions from equation~(\ref{eq:ps_w_cov_bin}) for the linear (dotted lines) and smoothed non-linear (dashed lines) input power spectra.
The upper and lower panels show the cases of two wedges, $\sigma_{P_{\perp,\parallel}}(k_i)$, and three wedges, $\sigma_{P_{3\mathrm{w},n}}(k_i)$, respectively.}
\label{fig:minerva_PS_w_cov_large}
\end{figure}

Assuming three PS wedges, we label the covariance between the $n$-th and $m$-th wedge as $C^P_{3\mathrm{w},n,m}(k_i,k_j)$ and the dispersion as $\sigma^P_{3\mathrm{w},n}(k_i) \equiv [C^P_{3\mathrm{w},n,n}(k_i,k_i)]^{1/2}$ for clarity.
For the case of two clustering wedges, we stick with the labels $\perp$ and $\parallel$.
Fig{.}~\ref{fig:minerva_PS_w_cov_small} compares the predicted covariance of two (upper panel) and three (lower panel) wedges to the data covariance measured from our 100 realizations for the small binning configuration.
The same comparison is plotted for the case of the large bins in Fig{.}~\ref{fig:minerva_PS_w_cov_large}.
In the linear-theory predictions, the BAO peaks and troughs in the dispersion are not as damped as in the data, especially for the most parallel wedges.
Further, using linear theory the high-$k$ dispersion of the intermediate and most perpendicular wedge is underpredicted ($k \gtrsim 0.15 \; h / \unit{Mpc}$).

\begin{figure}
 \includegraphics[width=\columnwidth,page=2]{images/minerva_ps_3wedges_cov_mat.pdf}
 \includegraphics[width=\columnwidth,page=1]{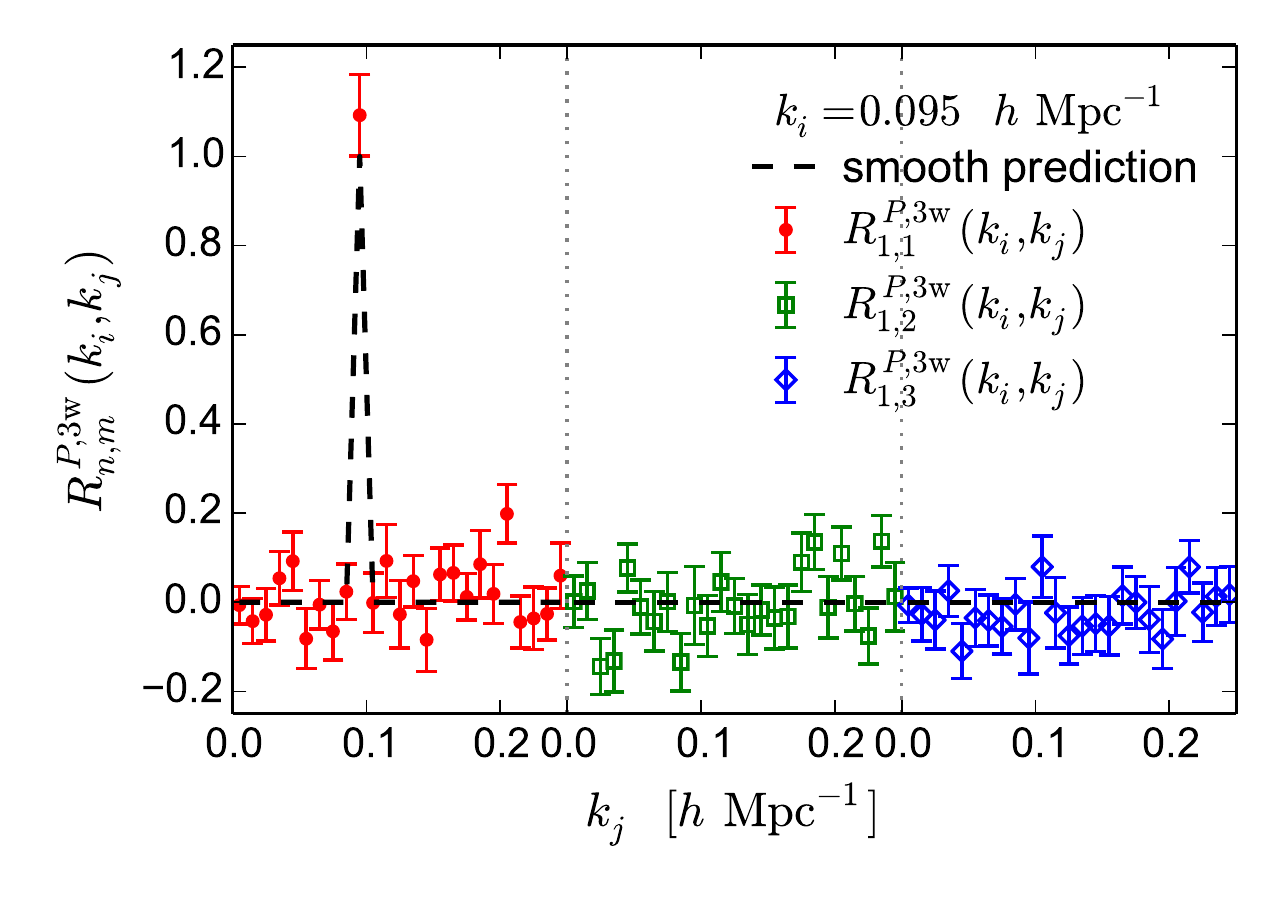}
 \caption{\emph{Upper panel:} The full correlation matrix of the PS wedges, $R_{3\mathrm{w},n,m}(k_i,k_j) = C^P_{3\mathrm{w},n,m}(k_i,k_j) [\sigma_{P^\mathrm{nl}_{3\mathrm{w},n}}(k_i) \, \sigma_{P^\mathrm{nl}_{3\mathrm{w},m}}(k_j)]^{-1}$, shows a dominant diagonal and no significant cross-correlation between wedges (shown here: large binning scheme).
  \emph{Lower panel:} Cut through the correlation matrix for the most transverse wedge, $\xi_{3\mathrm{w},1}$, at $k_i = 0.095 \; h \, \unit{Mpc}^{-1}$.
  The correlation contamination from physical effects not accounted for by our model and from noise is at the $15\%$ level (same level as the error bars).}
 \label{fig:minerva_PS_3w_cov_mat}
\end{figure}

We complete the PS wedge covariance analysis by showing the full correlation matrix compared to the non-linear prediction in Fig{.}~\ref{fig:minerva_PS_3w_cov_mat}.
Here we restrict the analysis to the 3-wedges case for brevity; the 2-wedges correlation matrix has similar properties.
As described in section~\ref{sec:cov_model}, the cross-correlation $C^P_{3\mathrm{w},n,m}(k_i, k_j)$ vanishes for $n \neq m$ or $k_i \neq k_j$.
The data correlation matrix, $R_{3\mathrm{w},n,m}(k_i,k_j) = C^P_{3\mathrm{w},n,m}(k_i,k_j) [\sigma_{P^\mathrm{nl}_{3\mathrm{w},n}}(k_i) \, \sigma_{P^\mathrm{nl}_{3\mathrm{w},m}}(k_j)]^{-1}$, again normalized by the non-linear theoretical prediction, shows a level of unaccounted cross-correlation of up to 15\%.
As for the multipoles, this is most likely noise contamination and the scatter cannot be associated with any systematic deviations from our Gaussian model.

\subsubsection{Configuration Space Covariance}

\begin{figure}
 \includegraphics[width=\columnwidth]{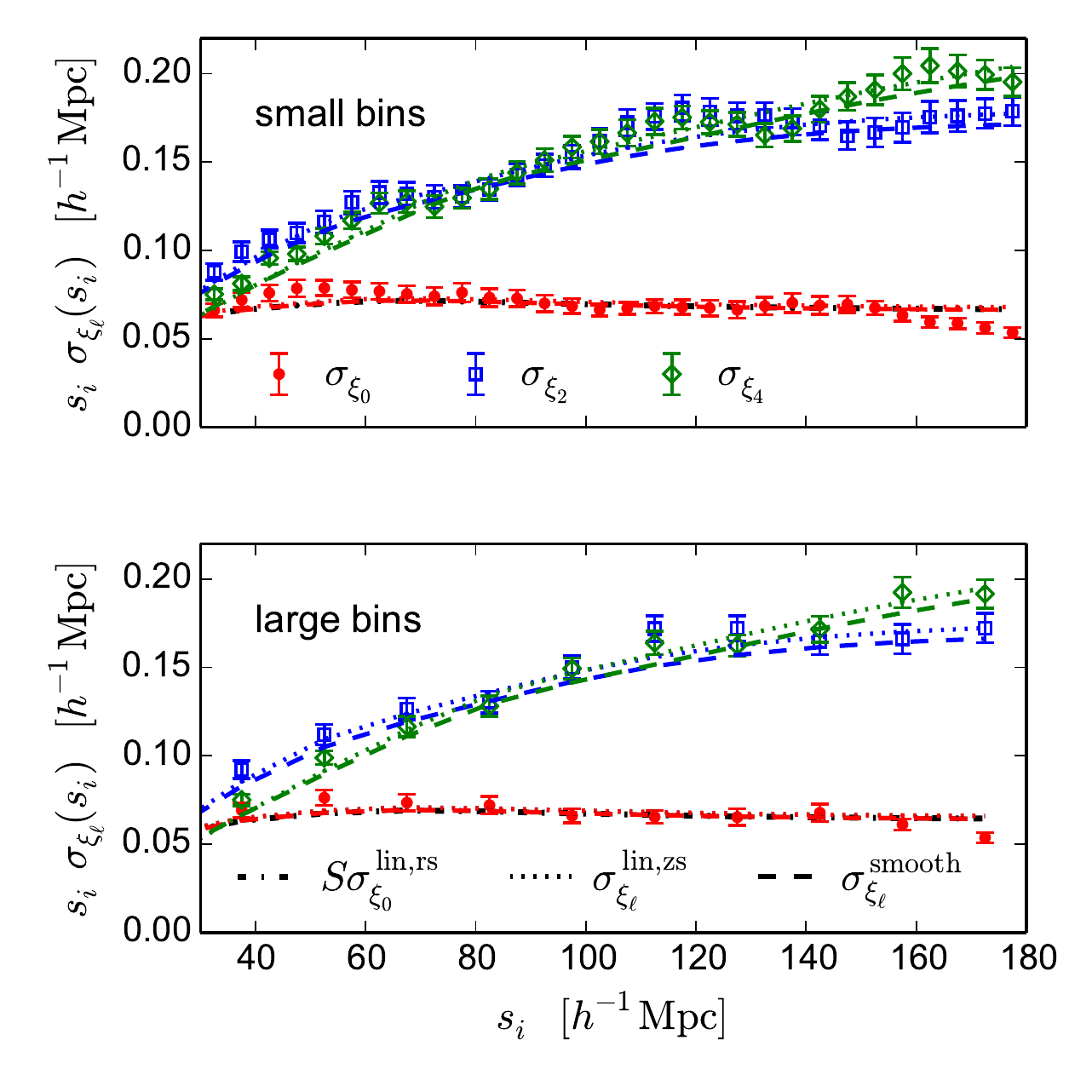}
 \caption{Comparison of the dispersion of the 2PCF multipoles, $\sigma_{\xi_\ell}(s_i)$ of our HOD realizations (points) compared with the Gaussian predictions from equation~(\ref{eq:2pcf_ell_cov_bin}) for the small (upper panel) and large (lower panel) binning configuration.
 The dotted and dashed lines correspond to the linear and smoothed non-linear input power spectra, respectively.
 In addition, we show the prediction of the real-space covariance as given in \citet{Sanchez:2008iw}, rescaled by the linear Kaiser factor $S = 1.28$ as dot-dashed black line.
 This curve cannot be distinguished from the Gaussian prediction for the monopole given in this work.}
 \label{fig:minerva_2PCF_ell_cov}
\end{figure}

In configuration space, we define the dispersion of 2PCF multipoles by $\sigma^\xi_\ell(s_i) \equiv [C^\xi_{\ell\ell}(s_i,s_i)]^{1/2}$.
In Fig{.}~\ref{fig:minerva_2PCF_ell_cov}, the predicted dispersion of the 2PCF multipoles is compared to the data dispersion measured from our 100 simulations for the small (upper panel) and large (lower panel) binning setup.
The difference between linear and non-linear predictions for the multipole covariance is negligible compared to the deviations between the measurements and the theory estimates.
Those deviations are largely due to the low number of realizations and the fact that the Fourier transform to configuration space translates uncorrelated modes into highly correlated measurements of the binned anisotropic 2PCF.
The dispersion of higher-order multipole moments of the correlation function is discussed in appendix~\ref{app:2pcf_multipoles}.

The correlation between measurement bins is best illustrated by the full information encoded in the correlation matrix, which is defined in an analogous way to the case of PS multipoles,
\begin{equation}
  R^\xi_{\ell_1\ell_2}(s_i,s_j) = C^\xi_{\ell_1\ell_2}(s_i,s_j) \, \left[ \sigma_{\xi^\mathrm{nl}_{\ell_1}}(s_i) \, \sigma_{\xi^\mathrm{nl}_{\ell_2}}(s_j) \right]^{-1}.
\end{equation}
Cross-covariance terms, $C^\xi_{\ell_1\ell_2}(s_i,s_j)$, show a large scatter due to the low signal-to-noise of our measurement.
We find that the differences in our linear and non-linear modelling are much smaller than the scatter in the measurements and we conclude agreement between the different predictions and data on an equally good level.

\begin{figure}
 \includegraphics[width=\columnwidth,page=2]{images/minerva_xi_multipoles_cov_mat.pdf}
 \includegraphics[width=\columnwidth,page=1]{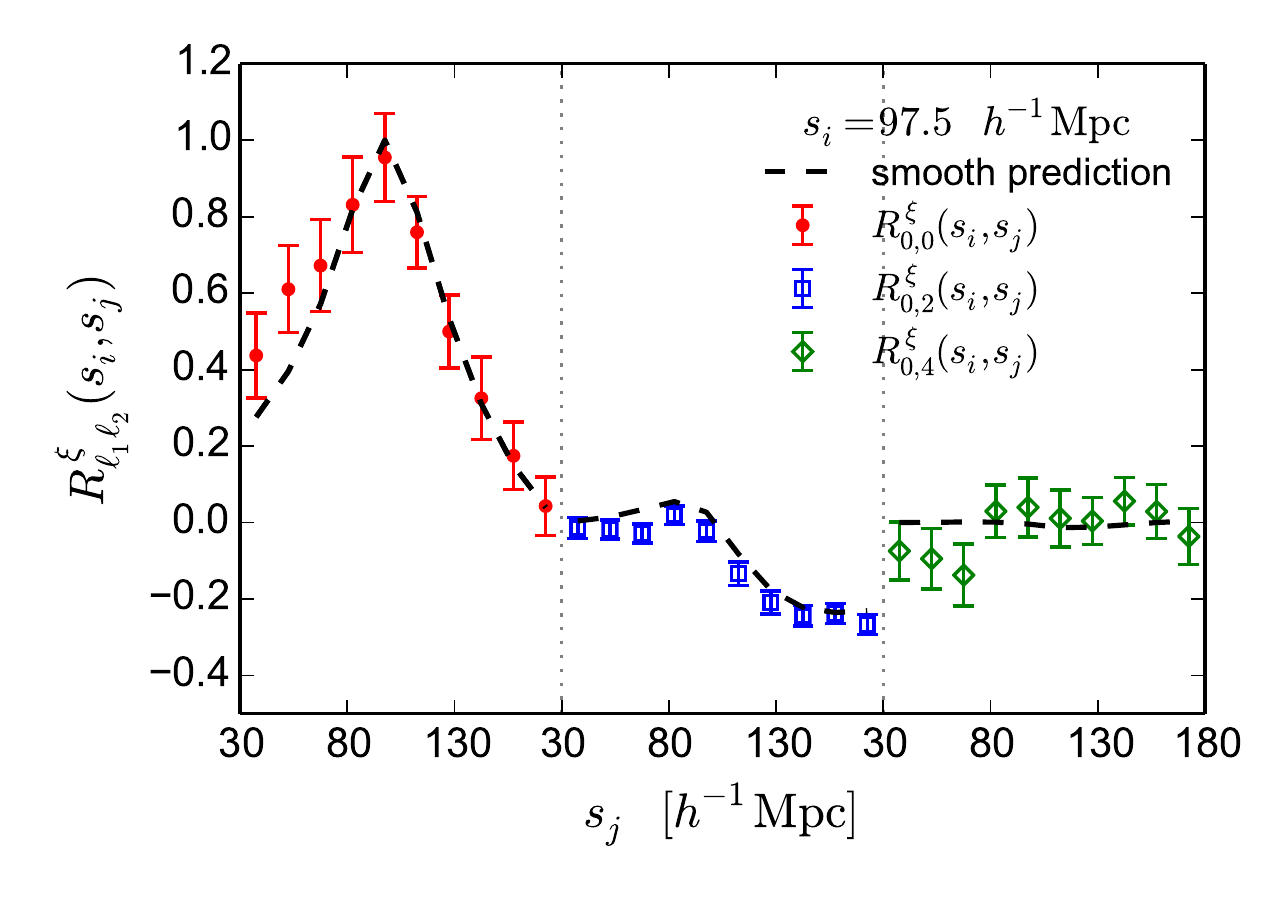}
 \caption{\emph{Upper panel:} In the full correlation matrix of the 2PCF multipoles, $R^\xi_{\ell_1\ell2}(s_i,s_j) = C^\xi_{\ell_1\ell_2}(s_i,s_j) [\sigma_{\xi^\mathrm{nl}_{\ell_1}}(s_i) \, \sigma_{\xi^\mathrm{nl}_{\ell_2}}(s_j)]^{-1}$ (normalized by the theoretical prediction), shown here for the large binning, we see a high level of cross-correlation which is only slowly declining away from the main diagonals and the diagonals of the monopole-quadrupole and quadrupole-hexadecapole sub-matrices.
 The monopole and the hexadecapole are not correlated.
  \emph{Lower panel:} Cut through the correlation matrix for $\ell_1 = 0$ at $s_i = 112.5 \; h^{-1} \, \unit{Mpc}$.
  Although the correlation curves inferred from our simulations are noisy, they follow the same trends as out our theoretical predictions.}
 \label{fig:minerva_2PCF_ell_cov_mat}
\end{figure}

In the upper panel of Fig{.}~\ref{fig:minerva_2PCF_ell_cov_mat}, we show the structure in the data correlation matrix (upper triangular part) next to the theoretical prediction (lower triangular part).
For a better visualization of the level of correlation, the lower panel shows a cut through the matrix.
The prediction for the monopole, quadrupole and hexadecapole correlation is in very good agreement with the measurements.
We see however that the data cross-correlations between different multipole moments (especially between monopole and hexadecapole) are noisy.

\begin{figure}
 \includegraphics[width=\columnwidth]{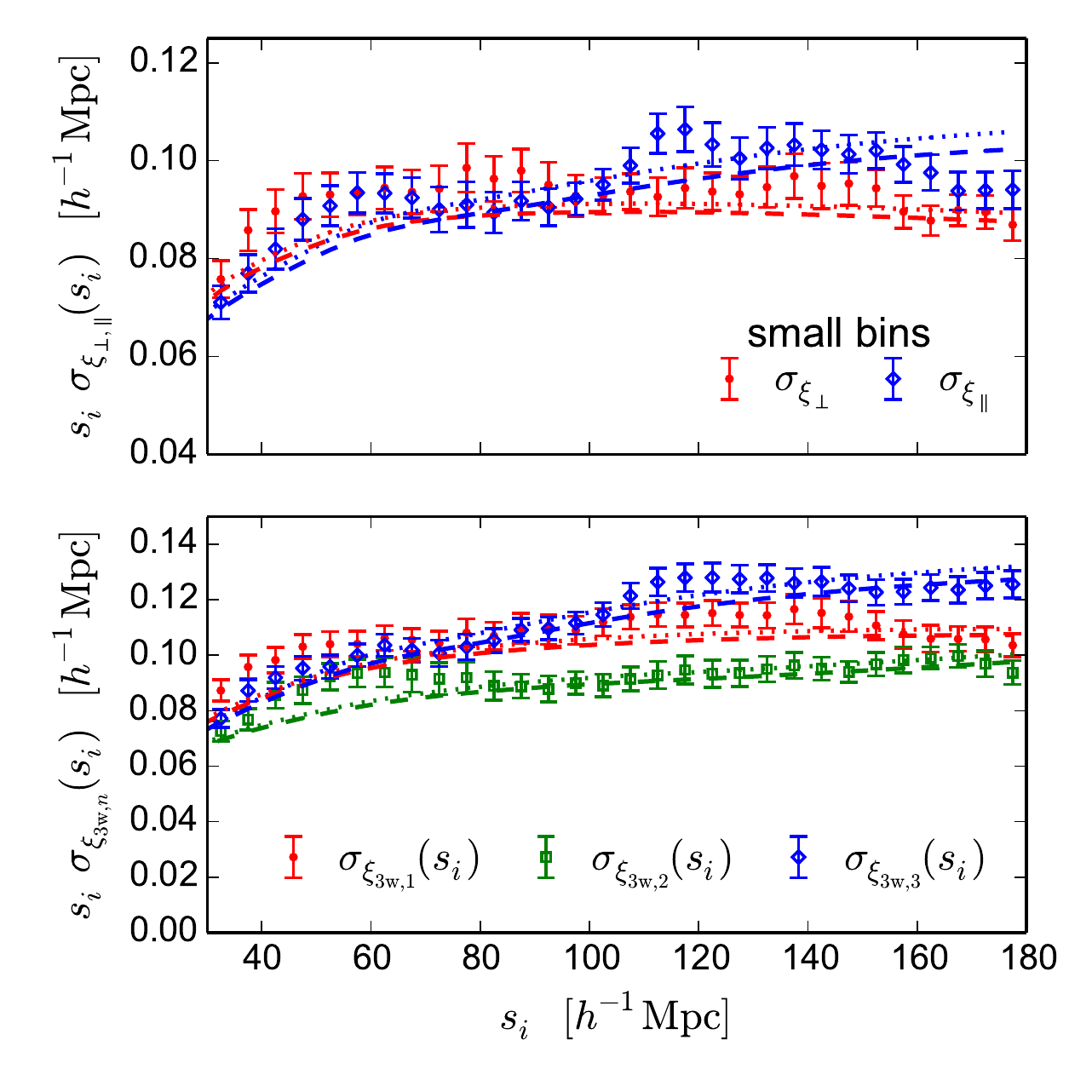}
 \caption{Comparison of the dispersion of the 2PCF wedges of our HOD realizations (points) for the small binning configuration with the Gaussian predictions from equation~(\ref{eq:2pcf_w_cov_bin}).
  The dotted and dashed lines correspond to the linear and smoothed non-linear input power spectra, respectively.
  The upper and lower panels show the cases of two wedges, $\sigma_{\xi_{\perp,\parallel}}(s_i)$, and three wedges, $\sigma_{\xi_{3\mathrm{w},n}}(s_i)$, respectively.}
 \label{fig:minerva_2PCF_w_cov_small}
\end{figure}

\begin{figure}
 \includegraphics[width=\columnwidth]{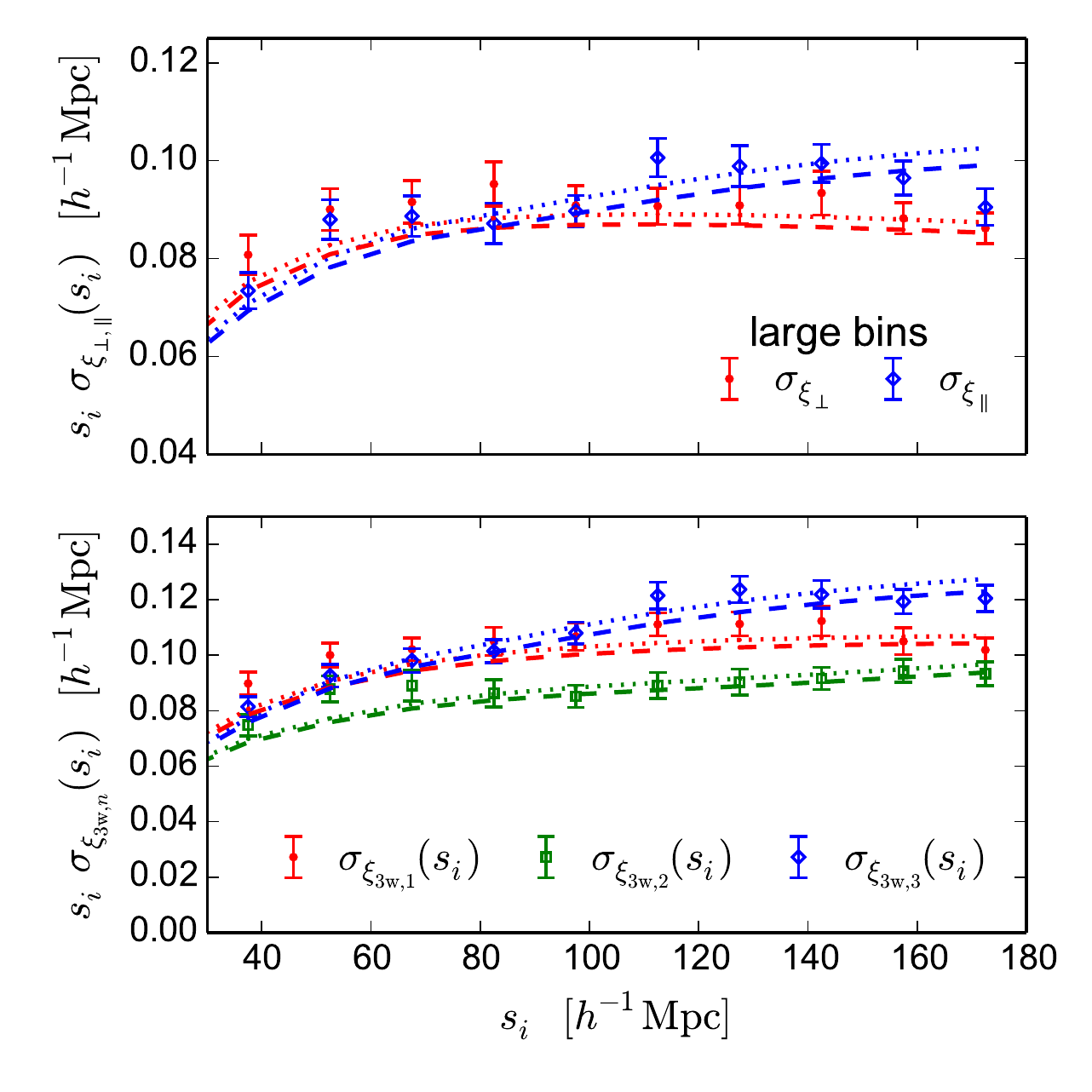}
 \caption{Comparison of the dispersion of the 2PCF wedges of our HOD realizations (points) for the large binning configuration with the Gaussian predictions from equation~(\ref{eq:2pcf_w_cov_bin}).
  The dotted and dashed lines correspond to the linear and smoothed non-linear input power spectra, respectively.
  The upper and lower panels show the cases of two wedges, $\sigma_{\xi_{\perp,\parallel}}(s_i)$, and three wedges, $\sigma_{\xi_{3\mathrm{w},n}}(s_i)$, respectively.}
 \label{fig:minerva_2PCF_w_cov_large}
\end{figure}

As for PS wedges, we label the covariance between the $n$-th and $m$-th 2PCF wedge as $C^\xi_{3\mathrm{w},n,m}(s_i,s_i)$ and the dispersion as $\sigma^\xi_{3\mathrm{w},n}(s_i) \equiv [C^\xi_{3\mathrm{w},n,n}(s_i,s_i)]^{1/2}$.
In the case of two wedges only, we label each one with a subscript $\perp$ or $\parallel$.

In Fig{.}~\ref{fig:minerva_2PCF_w_cov_small}, we plot the predicted dispersion of two (upper panel) and three (lower panel) wedge measurements in configuration space compared to the data dispersion measured from the \Minerva simulation for the small binning configuration.
These predictions are based on equation~(\ref{eq:2pcf_w_cov_bin}), including contributions up to $\ell_1, \ell_2 \le 6$.
We discuss the convergence of the wedge dispersion with the inclusion of higher-order multipoles in appendix~\ref{app:2pcf_wedges}.
The same comparison is plotted for the case of the large bins in Fig{.}~\ref{fig:minerva_2PCF_w_cov_large}.
In both cases, there is no significant difference between the linear and the non-linear predictions.
The estimate from linear theory is always slightly larger than the estimate from the smoother interpolation of the measured power spectrum multipoles, but this difference is much smaller than the deviation between theory and data.

\begin{figure}
 \includegraphics[width=\columnwidth,page=2]{images/minerva_xi_3wedges_cov_mat_maxl6.pdf}
 \includegraphics[width=\columnwidth,page=1]{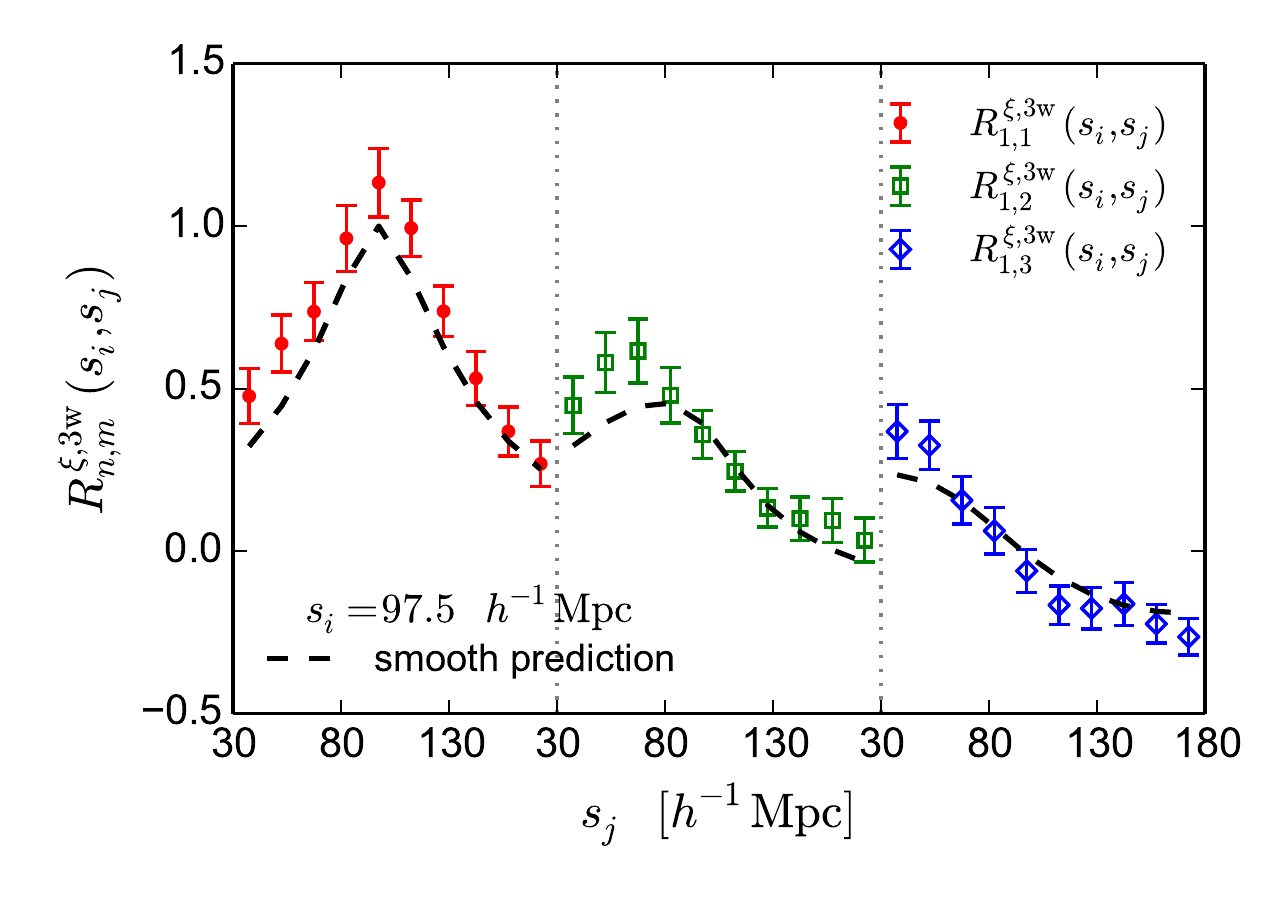}
 \caption{\emph{Upper panel:} The full correlation matrix of the 2PCF wedges, $\corr{\xi_{3\mathrm{w},1}(k_i),\xi_{3\mathrm{w},n}(k_j)}$, is dominated by a high level of correlation between different wedges and distance bins (shown here for the large binning configuration).
  \emph{Lower panel:} Cut through the correlation matrix for the most transverse wedge, $\xi_{3\mathrm{w},1}$, at $s_i = 112.5 \; h^{-1} \, \unit{Mpc}$.
  Although the data cross-correlations are noisy, they are well described by our theoretical prediction (the significant correlation of the covariance is discussed in the text).}
 \label{fig:minerva_2PCF_3w_cov_mat}
\end{figure}

The cross-correlation for 2PCF wedges is as significant as for the configuration-space multipoles.
In analogy to the discussion on the Fourier space wedges, we define the correlation matrix for 2PCF wedges as $R^\xi_{3\mathrm{w},n}(s_i,s_j) = C^\xi_{3\mathrm{w},n,m}(s_i,s_j) [\sigma_{\xi^\mathrm{nl}_{3\mathrm{w},n}}(s_i) \, \sigma_{\xi^\mathrm{nl}_{3\mathrm{w},m}}(s_j)]^{-1}$, normalized by the non-linear theoretical prediction.
We show the full correlation matrix compared to the non-linear prediction in Fig{.}~\ref{fig:minerva_2PCF_3w_cov_mat}.
Again, we restrict our discussion to the case of three wedges for brevity.
The correlation matrix for two 2PCF wedges has similar properties.
As the Fourier transformation mixes independent Fourier modes, the covariance $C^\xi_{3\mathrm{w},n,m}(s_i, s_j)$ only decays slowly with increased separation between the distance bins and ranges of the LOS parameter.
The data correlation (upper panel of Fig{.}~\ref{fig:minerva_2PCF_3w_cov_mat}) shows good overall agreement between the theoretical prediction (lower triangular part) and our measurement (upper triangular part).
The plotted full matrix shows also that the cross-correlation structure for 2PCF wedges is more complex than for 2PCF multipoles.
For better visualization, we show a cut through the matrix in the lower panel of Fig.~\ref{fig:minerva_2PCF_3w_cov_mat}.
We find differences between our measurements and the theoretical predictions of up to 20\%, which is a bit higher than for the multipole case.
As for the previous cases, these discrepancies are most likely due to noise contamination and the level of noise is very similar for different cuts through the matrix.
The model under- or overpredicts the data over wide ranges of pair separations because the noise in the covariance is as correlated as the 2PCF itself.
Hence, we are not able to test for deviations from of our Gaussian ansatz with the number of realizations at hand.

As already discussed in section~\ref{sec:cov_model}, the shortcomings of our model for the analysed sample of N-body simulations is the neglected trispectrum contribution which is only relevant for scales in the non-linear regime.
We would need a much larger set of N-body simulations in order to validate extensions to the Gaussian model.

As a final test, we analyse whether our model for the monopole covariance has improved over the simplifying assumptions used by \citet{Sanchez:2008iw}, where the Gaussian real-space covariance from linear theory has been rescaled by the Kaiser factor for linear RSD.
This model defines the monopole covariance in real-space as follows:
\begin{multline}
 C^{\xi,\mathrm{lin,rs}}_{0,0}(s_i, s_j) = \frac{1}{\pi^2 \, V_\mathrm{s}} \int_0^\infty \left[ \bgal^2 \, P_\mathrm{L}(k) + \frac{1}{\bar n} \right]^2 \\
 \times \jlbar(ks_i) \, \jlbar(ks_j) \, k^2 \dint k,
\end{multline}
From this, the $z$-space covariance is estimated to be $C^{\xi,\mathrm{lin,zs}}_{0,0}(s_i, s_j) = S^2 \, C^{\xi,\mathrm{lin,rs}}_{0,0}(s_i, s_j)$, where $S = 1 + \frac 2 3 \beta + \frac 1 5 \beta^2$.
As shown in Fig{.}~\ref{fig:minerva_2PCF_ell_cov},
this simple ansatz gives already an excellent description of the covariance of the redshift-space 2PCF monopole, even though the anisotropy of the input power spectrum has been neglected.
Any RSD or BAO fit using anisotropic clustering measurements must take the higher-order multipoles into account which can only be done by the formulae presented in this work.

Additional tests indicate that the discrepancy between the simplified monopole recipe and the full Gaussian model presented here depends on the bias.
While there is a negligible difference for the highly biased HOD galaxy sample analysed here (where $b^2 = 4.02$), the covariance of an under-biased tracer sample ($b \lesssim 1$, similar $\bar n$) is only correctly modelled taking the full anisotropy of the input power spectrum into account.

%%%%% CONCLUSIONS %%%%%

\section{Conclusions}
\label{eq:conclusions}

In this work, we presented explicit formulae for the Gaussian covariance matrix of anisotropic two-point clustering measurements.
Aiming at precise covariance estimates for the verification of models of the redshift-space clustering two-point statistics, we looked at the covariance of clustering wedges (\ie, large bins in the line-of-sight parameter) and multipole moments of the clustering signal, both in configuration space as well as in Fourier space.
The formulae presented here rely on a model for the input power spectrum.
In our analysis, we tested two different prescriptions for $P(k,\mu)$: first, a linear model for the redshift-space galaxy power spectrum
based on the linear Kaiser effect and a linear galaxy bias on top of the linear-theory prediction for the matter power spectrum.
Second, we use a smoothed interpolation of the measured galaxy power spectrum from our simulations.

The covariance model has been tested and validated against a set of large-volume N-body simulations.
The number of realizations is enough to have a precise mean measurement of the non-linear anisotropic clustering signal, but the recovered covariance 
matrices exhibit significant noise due to the relatively small number of realizations.

We show that the Gaussian covariance from the interpolated non-linear power spectrum describes the measured clustering covariance very well.
Not only is the overall shape of the dispersion of PS wedges and multipoles in excellent agreement, also broad features such as damped BAOs are described accurately.
Relying on the linear power spectrum model, the Gaussian prediction is also in very good agreement with the observed data covariance, but the BAO features in the PS covariance are slightly overpredicted.
In configuration space, the measurements of the anisotropic clustering are highly correlated.
The theory predictions for the covariance of wedges and multipoles of the two-point correlation function (2PCF) accurately describe the full correlation structure.
The differences in these covariances resulting from the two different descriptions for the input power spectrum are almost negligible.

The aim of this work is limited to predictions of the anisotropic clustering covariance on quasi-linear scales in order to allow RSD and BAO fits of the clustering measurements from a limited set of large-volume N-body realizations.
Further work is needed to incorporate the beat-coupling with super-survey modes and the contributions from the connected trispectrum part of the covariance with arises due to non-linear evolution \citep{Scoccimarro:1999kp}.
These effects have been neglected here due to absence of super-survey modes (super sample covariance; SSC) in the N-body realizations and the restrictions to quasi-linear scales ($k \lesssim 0.2 \; h \, \mathrm{Mpc}^{-1}$, $s \gtrsim 40 \; h^{-1} \, \unit{Mpc}$).
In order to validate a model for the influence of SSC on the anisotropic clustering covariance, an similar analysis to the work in  \citet{dePutter:2011ah,Takada:2013wfa,Li:2014sga} -- \ie, taking account of the local density estimate and the beat-coupling of smaller scales with super-survey modes -- is required, which is beyond the scope of the work presented here.

We expect careful modelling to be more important the more complex the analysed galaxy sample is, and the deeper in the non-linear regime the clustering probe advances.
The Gaussian model derived here can easily be extended to account for the dependence of the covariance on cosmological parameters by varying the input power spectrum.
In future work, we plan to extend our studies to models of the clustering covariance of surveys with non-trivial geometry -- for the analytical treatment in case of isotropic clustering measurements, see \citet{dePutter:2011ah}.
After submission of our initial manuscript, \citet{OConnell:2015mwa} published a pre-print presenting a way to take into account non-Gaussian contributions to the anisotropic 2PCF covariance matrix for a CMASS-like survey.

As final remark, we want to emphasize that the set of \Minerva simulations is especially suited for RSD model testing in the mildly non-linear regime and such tests can make use of the covariance matrices presented in this work.
Among other tests, we will apply this in the context of the RSD analysis of the final release of the BOSS galaxy clustering sample currently being prepared.

\section*{Acknowledgements}

We thank Daniel Farrow, Elisabeth Krause, Martha Lippich, and Francesco Montesano for helpful discussions.
We would like to thank the anonymous reviewer of our initial submission for many helpful comments.
JNG, AGS and SSA acknowledge support from the Transregional Collaborative Research Centre TR33 `The Dark Universe' of the German Research Foundation (DFG).
CDV acknowledges support from the Ministry of Economy and Competitiveness (MINECO, Spain) through grants AYA2013-46886 and AYA2014-58308.
The analysis has been performed on MPE's computing cluster for the \emph{Euclid} project and the \emph{Hydra} cluster at the Rechenzentrum Garching (RZG).

%%%%%%%%%%%%%%%%%%%%%%%%%%%%%%%%%%%%%%%%%%%%%%%%%%


\bibliographystyle{mnras}
\bibliography{covpaper}

%%%%%%%%%%%%%%%%%%%%%%%%%%%%%%%%%%%%%%%%%%%%%%%%%%

%%%%%%%%%%%%%%%%% APPENDICES %%%%%%%%%%%%%%%%%%%%%

\onecolumn

\appendix

\section{Proofs of the Binned Covariance Relations}
\label{app:cov_proofs}
In this section, we present the full derivation of the various formulae given in section~\ref{sec:cov_model} for the covariance of binned anisotropic clustering measurements.

We assume $\Nbins$ wavenumber bins centred at $k_i$ with width $\Delta k$.
In the following we will average equation~(\ref{eq:ps_cov}) over such bins, which have a volume $V_{k_i} = 4 \pi [(k_i + \Delta k/2)^3 - (k_i - \Delta k/2)^3]/3$.
This is easily performed by integration in spherical coordinates, so that $\delta_\mathrm{D}(\V k - \VPrime k) = \delta_\mathrm{D}(k - k') \delta_\mathrm{D}(\theta - \theta') \delta_\mathrm{D}(\phi - \phi') / (2 \pi k^2 \sin \theta)$, and we will make use of the following:
\begin{multline}
 \label{eq:ps_cov_int}
 \frac{1}{V_{k_i}} \int_{V_{k_i}} \dnx{3}{k} \frac{1}{V_{k_j}} \int_{V_{k_j}} \dnx{3}{k'} \Cov{P(\V k), P(\VPrime k)} = \frac{2 \, (2 \pi)^5}{V_{k_i} \, V_{k_j} \, V_\mathrm{s}} \int_{k_i-\Delta k/2}^{k_i+\Delta k/2} k^2 \dint k \int_{k_j-\Delta k/2}^{k_j+\Delta k/2} (k')^2 \dint k' \\
 \times \int_{-\pi/2}^{\pi/2} \sin(\theta) \dint \theta \int_{-\pi/2}^{\pi/2} \sin(\theta') \dint \theta' \frac{\delta_\mathrm{D}(k - k') \delta_\mathrm{D}(\theta - \theta')}{2 \pi k^2 \sin (\theta)} \left[ P(k, \cos(\theta)) + \frac{1}{\bar n} \right]^2,
\end{multline}
where $V_\mathrm{s}$ is the volume of the galaxy sample.

\subsection{PS Multipoles}
\label{app:ps_multipoles}

\begin{figure}
 \centering
 \includegraphics[width=.43\columnwidth,trim=0 0 0 250,clip=true]{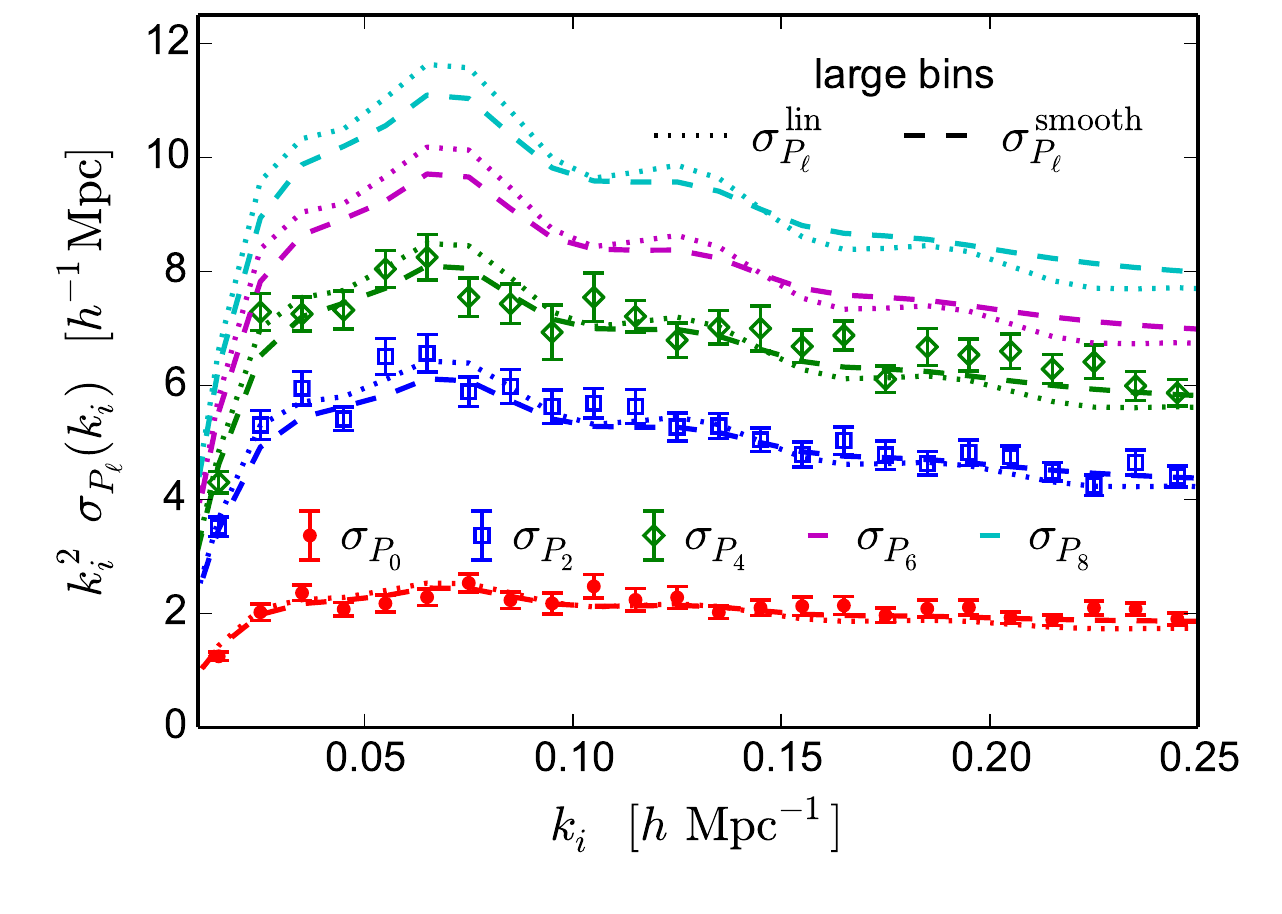}
 \includegraphics[width=.45\columnwidth,trim=0 0 0 250,clip=true]{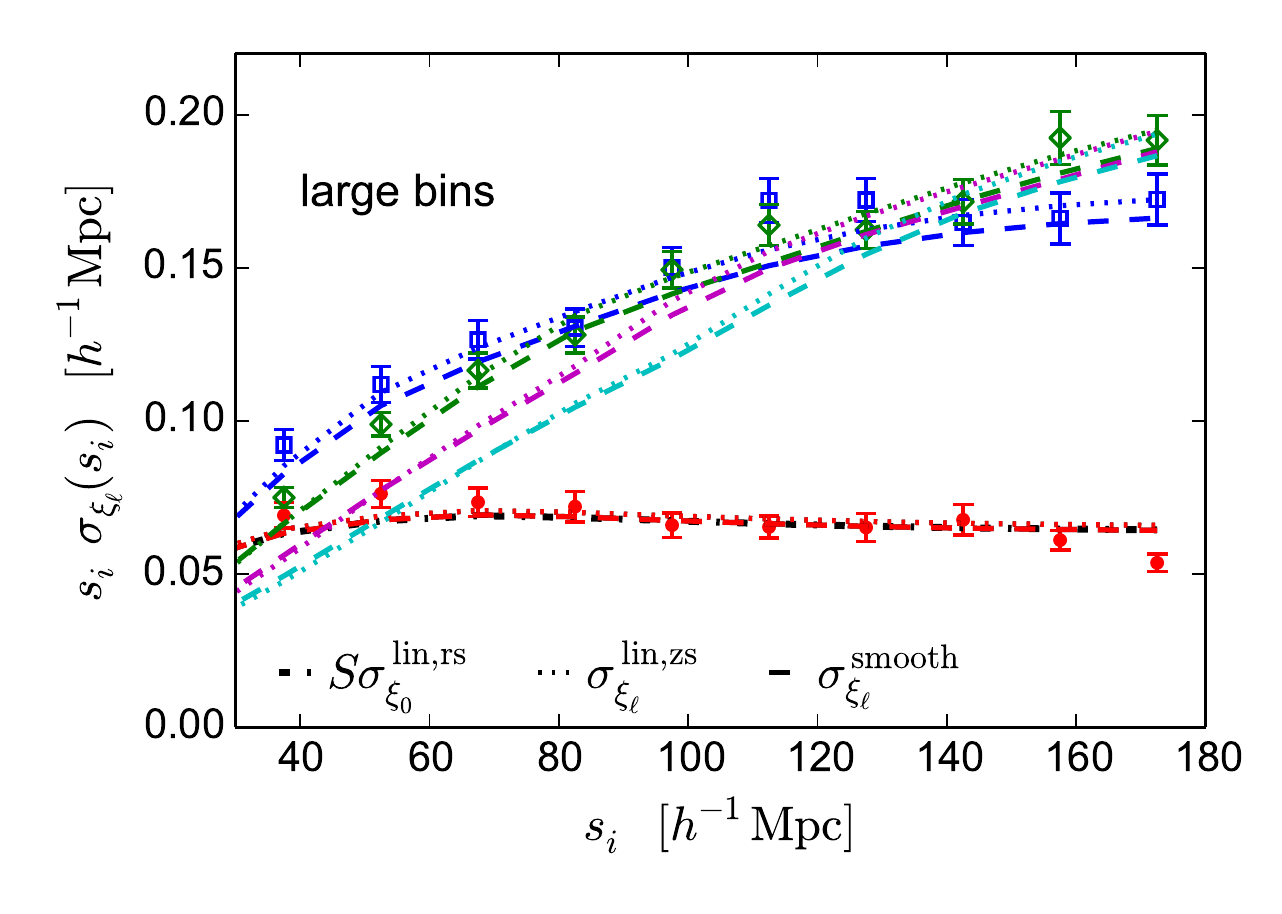}
 \caption{\emph{Left panel}: Same comparison of the PS multipoles dispersion as in the lower panel of Fig.~\ref{fig:minerva_PS_ell_cov} but including also the predictions for $\sigma_{P_6}$ and $\sigma_{P_8}$ for comparison.
  \emph{Right panel}: Likewise inclusion of the predictions for $\sigma_{\xi_6}$ and $\sigma_{\xi_8}$ to the comparison plot in the lower panel of Fig.~\ref{fig:minerva_2PCF_ell_cov}. For brevity, only the results for the large binning scheme are shown here.}
 \label{fig:minerva_ell_cov_moretheo}
\end{figure}

Replacing $P(k,\mu)$ by its multipole expansion, $P(k,\mu) = \sum_\ell P_\ell(k) \Lp(\mu)$, in the definition of the multipole covariance in equation~(\ref{eq:ps_cov_ell_ell}), we find
\begin{equation}
 \sigma^2_{\ell_1\ell_2} = \frac{(2 \ell_1 + 1) \, (2 \ell_2 + 1)}{V_\mathrm{s}} \sum_{\ell_3=0}^\infty \sum_{\ell_4=0}^{\ell_3} \left[ P_{\ell_4}(k) + \frac{1}{\bar n} \delta_{\ell_4 0} \right] \left[ P_{\ell_3-\ell_4}(k) + \frac{1}{\bar n} \delta_{(\ell_3-\ell_4) 0} \right] \int_{-1}^1 \Lp[\ell_1](\mu) \, \Lp[\ell_2](\mu) \, \Lp[\ell_4](\mu) \, \Lp[\ell_3-\ell_4](\mu) \dint \mu.
\end{equation}
At this point, we use the expansion of a product of two Legendre polynomials by use of Wigner $3j$-symbols \citep{rotenberg19593},
\begin{equation}
 \Lp[\ell_1](\mu) \Lp[\ell_2](\mu) = \sum_{\ell=|\ell_1-\ell_2|}^{\ell_1+\ell_2} \begin{pmatrix} \ell_1 & \ell_1 & \ell \\ 0 & 0 & 0 \end{pmatrix}^2 (2 \ell + 1) \Lp(\mu),
\end{equation}
and the orthogonality of the $\Lp(x)$, yielding
\begin{equation}
 \int_{-1}^1 \Lp[\ell_1](\mu) \Lp[\ell_2](\mu) \Lp[\ell_4](\mu) \Lp[\ell_3-\ell_4](\mu) \dint \mu = 2 \sum_{\ell=\max(|\ell_1-\ell_2|,|2\ell_4-\ell_3|)}^{\min(\ell_1+\ell_2,\ell_3)} \begin{pmatrix} \ell_1 & \ell_2 & \ell \\ 0 & 0 & 0 \end{pmatrix}^2 \begin{pmatrix} \ell_4 & \ell_3 - \ell_4 & \ell \\ 0 & 0 & 0 \end{pmatrix}^2.
\end{equation}
Inserting this back into the expression for $\sigma^2_{\ell_1\ell_2}$ gives
\begin{multline}
 \label{eq:ps_cov_ell_ell_in_ps_ell}
 \sigma^2_{\ell_1\ell_2}(k) = \frac{2 \, (2 \ell_1 + 1) \, (2 \ell_2 + 1)}{V_\mathrm{s}} \sum_{\ell_3=0}^\infty \sum_{\ell_4=0}^{\ell_3} \left[ P_{\ell_4}(k) + \frac{1}{\bar n} \delta_{\ell_4 0} \right] \left[ P_{\ell_3 - \ell_4}(k) + \frac{1}{\bar n} \delta_{(\ell_4 - \ell_3) 0} \right] \\
 \times \sum_{\ell=\max(|\ell_1-\ell_2|,|2\ell_4-\ell_3|)}^{\min(\ell_1+\ell_2,\ell_3)} \begin{pmatrix} \ell_1 & \ell_2 & \ell \\ 0 & 0 & 0 \end{pmatrix}^2 \begin{pmatrix} \ell_4 & \ell_3 - \ell_4 & \ell \\ 0 & 0 & 0 \end{pmatrix}^2,
\end{multline}
where $\delta_{\ell0}$ reflects that the shot noise contributes only to the monopole term.

First, we consider PS Legendre moments measured in the wavenumber bins defined above.
Their theoretical covariance matrix $C_{{\ell_1}{\ell_2}ij}$ can be obtained by the following integration over the bins:
\begin{equation}
 \label{eq:ps_ell_cov_int}
 \Cov{P^i_{\ell_1}, P^j_{\ell_2}} = \frac{(2 \ell_1 + 1) \, (2 \ell_2 + 1)}{V_{k_i} \, V_{k_j}} \int_{V_{k_i}} \int_{V_{k_j}} \Cov{P(\V k), P(\VPrime k)} \Lp[\ell_1](\cos(\theta)) \, \Lp[\ell_2](\cos(\theta')) \, \dnx{3}{k'} \, \dnx{3}{k}.
\end{equation}
This equation expressed in spherical coordinates as in equation~(\ref{eq:ps_cov_int}) reads
\begin{equation}
 \Cov{P^i_{\ell_1}, P^j_{\ell_2}} = \frac{2 \, (2 \ell_1 + 1) \, (2 \ell_2 + 1) \, (2 \pi)^4}{V_{k_i} V_{k_j} \, V_\mathrm{s}} \int_{k_i-\Delta k/2}^{k_i+\Delta k/2} k^2 \dint k  \int_{-1}^1 \dint \mu \left[ P(k, \mu) + \frac{1}{\bar n} \right]^2 \Lp[\ell_1](\mu) \, \Lp[\ell_2](\mu),
\end{equation}
where we already evaluated the Dirac delta functions.
Using the definition of $\sigma^2_{\ell_1\ell_2}$, this yields
\begin{equation}
 \Cov{P^i_{\ell_1}, P^j_{\ell_2}} = \frac{2 \, (2 \pi)^4}{V_{k_i}^2} \delta_{ij} \int_{k_i-\Delta k/2}^{k_i+\Delta k/2} \sigma^2_{\ell_1\ell_2}(k) \, k^2 \dint k,
\end{equation}
which corresponds to the result presented in equation~(\ref{eq:ps_ell_cov_bin}).

In particular, the monopole covariance will be given by
\begin{align}
 \Cov{P^i_0, P^j_0} &= \frac{2 \, (2 \pi)^4}{V_{k_i}^2 \, V_\mathrm{s}} \delta_{ij} \sum_{\ell_3=0}^\infty \sum_{\ell_4=0}^{\ell_3} \int_{k_i-\Delta k/2}^{k_i+\Delta k/2} k^2 \left[ P_{\ell_4}(k) + \frac{1}{\bar n} \delta_{\ell_4 0} \right] \left[ P_{\ell_3 - \ell_4}(k) + \frac{1}{\bar n} \delta_{(\ell_3 - \ell_4) 0} \right] \dint k \int_{-1}^1 \Lp[\ell_4](\mu) \, \Lp[\ell_3 - \ell_4](\mu) \dint \mu \\
 &= \frac{4 \, (2 \pi)^4}{V_{k_i}^2 \, V_\mathrm{s}} \delta_{ij} \sum_\ell \frac{1}{2 \ell + 1} \int_{k_i-\Delta k/2}^{k_i+\Delta k/2} k^2 \left[ P_\ell(k) + \frac{1}{\bar n} \delta_{\ell0} \right]^2 \dint k,
\end{align}
where we made use of the orthogonality of the $\Lp(\mu)$.
This relation could also have been derived by setting $\ell_1 = \ell_2 = 0$ in the Wigner $3j$-symbols in equation~(\ref{eq:ps_ell_cov_bin}).
This shows that the naive guess that monopole covariance would only be given by the monopole power spectrum is not correct.
Instead, the monopole covariance has contributions from all higher-order multipoles.

Note that, as implied by equation~(\ref{eq:ps_ell_cov_bin}), the covariance of higher-order multipoles does not vanish even though the input power spectrum is only modelled up to $\ell = 4$.
As shown in the left panel of Fig{.}~\ref{fig:minerva_ell_cov_moretheo}, the dispersion $\sigma_{P_\ell}$ increases with larger $\ell$.

\subsection{PS Clustering Wedges}
\label{app:ps_wedges}

For the theoretical covariance matrix $C_{\mu\mu'ij}$ of the PS clustering wedges, the integration over the wavenumber bins can be written as
\begin{align}
 \label{eq:ps_w_cov_ansatz}
 \Cov{P^i_\mu, P^j_{\mu'}} &= \frac{1}{V_{k_i} \, V_{k_j} \, (\Delta \mu)^2} \int_{V_{k_i},\mu \le \abs{\tilde \mu} \le \mu + \Delta \mu} \int_{V_{k_j},\mu' \le \abs{\tilde \mu'} \le \mu' + \Delta \mu} \Cov{P(\V k), P(\VPrime k)} \dnx{3}{k'} \dnx{3}{k} \\
 &= \frac{(2 \pi)^2}{V_{k_i} \, V_{k_j} \, (\Delta \mu)^2} \int_{k_i-\Delta k/2}^{k_i+\Delta k/2} k^2 \int_{k_j-\Delta k/2}^{k_j+\Delta k/2} (k')^2 \int_{\mu \le \abs{\tilde \mu} \le \mu + \Delta \mu} \int_{\mu' \le \abs{\tilde \mu'} \le \mu' + \Delta \mu} \Cov{P(\V k), P(\VPrime k)} \dint \tilde \mu \dint \tilde \mu' \dint k' \dint k
\end{align}
by use of a similar integration as in equation~(\ref{eq:ps_cov_int}) restricted to the volumes of each wedge, $\Delta \mu V_{k_i}$.
The evaluation of the Dirac delta functions in $\cov{P(\V k), P(\VPrime k)}$ yields
\begin{equation}
 \Cov{P^i_\mu, P^j_{\mu'}} = \frac{4 \, (2 \pi)^4}{V_{k_i}^2 \, V_\mathrm{s} \, (\Delta \mu)^2} \delta_{ij} \, \delta_{\mu \mu'} \int_{k_i-\Delta k/2}^{k_i+\Delta k/2} k^2 \dint k \int_\mu^{\mu + \Delta \mu} \left[ P(k, \tilde \mu) + \frac{1}{\bar n} \right]^2 \dint \tilde \mu,
\end{equation}
We already presented this result in equation~(\ref{eq:ps_w_cov_bin}).
An additional factor of 2 comes from the fact that the symmetry in $\mu \to -\mu$ has been used to simplify the integration over the $\mu$ range into a single contiguous interval.

\subsection{2PCF Multipoles}
\label{app:2pcf_multipoles}

Let us first consider the covariance of the (unbinned) 2PCF multipole moments which we will derive from the results in Fourier space by use of equation~(\ref{eq:xi_ell_from_ps}):
\begin{equation}
 \label{eq:2pcf_ell_cov}
 \Cov{\xi_{\ell_1}(\V s), \xi_{\ell_2}(\VPrime s)} = \frac{1}{(2 \pi)^6} \int \dnx{3}{k} \int \dnx{3}{k'} \Cov{P_{\ell_1}(\V k), P_{\ell_2}(\VPrime k)} \, \ft{\V k}{\V s}  \, \ft{\VPrime k}{\VPrime s},
\end{equation}
where the covariance of the power spectrum multipoles is given by
\begin{equation}
 \Cov{P_{\ell_1}(\V k), P_{\ell_2}(\VPrime k)} = (2 \pi)^3 \delta_\mathrm{D}(\V k - \VPrime k) \, \sigma^2_{\ell_1\ell_2}(k).
\end{equation}

The resulting expression for the covariance of 2PCF multipoles,
\begin{equation}
 \Cov{\xi_{\ell_1}(s), \xi_{\ell_2}(s')} = \frac{1}{(2 \pi)^3} \int \sigma^2_{\ell_1 \ell_2}(k) \, \ft{\V k}{\V s} \, \ft{\V k}{\VPrime s} \, \dnx{3}{k},
\end{equation}
can be simplified by use of the same transformation as in equation~(\ref{eq:xi_ell_from_ps_1d}),
\begin{equation}
 \Cov{\xi_{\ell_1}(s), \xi_{\ell_2}(s')} = \frac{\ii^{\ell_1+\ell_2}}{2 \pi^2} \int_0^\infty \sigma^2_{\ell_1\ell_2}(k) \, \jl[\ell_1](ks) \, \jl[\ell_2](ks') \, k^2 \, \dint k.
\end{equation}
This equation was already given in an almost identical form by \citet{White:2014naa}, but the authors did not explicitly indicate the shot-noise contribution that enters the variance per power spectrum mode, $\sigma^2_{\ell_1\ell_2}(k)$.

The binned covariance of 2PCF multipole moments is then given by the volume average of the spherical Bessel functions over the distance bins as already presented in equation~(\ref{eq:2pcf_ell_cov_bin}),
\begin{equation}
 \Cov{\xi^i_{\ell_1}, \xi^j_{\ell_2}} = \frac{\ii^{\ell_1+\ell_2}}{2 \pi^2} \int_0^\infty k^2 \, \sigma^2_{\ell_1\ell_2}(k) \, \jlbar[\ell_1](ks_i) \, \jlbar[\ell_2](ks_j) \dint k,
\end{equation}
where we made use of the spherical Bessel functions average over the distance bin around $s_i$ with volume $V_{s_i} = 4 \pi \left( s^3_{i,\mathrm{max}} - s^3_{i,\mathrm{min}} \right) / 3$ defined by
\begin{equation}
 \label{eq:jlbar}
 \jlbar(ks_i) \equiv \frac{4 \pi}{V_{s_i}} \int_{s_i - \Delta s/2}^{s_i + \Delta s/2} s^2 \, \jl(ks) \dint s.
\end{equation}
In contrast to the power spectrum case, the dispersion of higher-order multipole moments of the correlation function does not increase with $\ell$.
In fact, we show in the right panel of Fig{.}~\ref{fig:minerva_ell_cov_moretheo} that $\sigma_{\xi_6}$ and $\sigma_{\xi_8}$ follow a very similar trend than the hexadecapole deviation.

\subsection{2PCF Clustering Wedges}
\label{app:2pcf_wedges}

\begin{figure}
 \centering
 \includegraphics[width=.9\columnwidth]{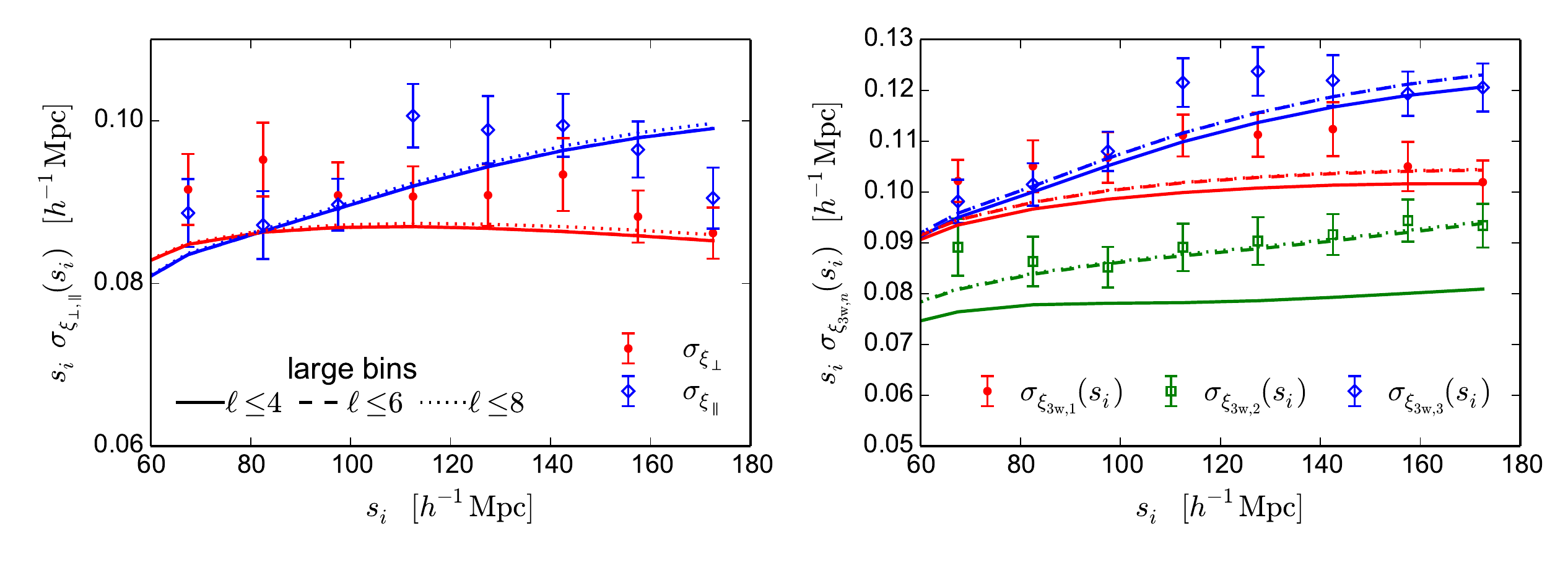}
 \caption{Convergence of the dispersion of two configuration space wedges (left panel) and three wedges (right panel) with the multipole order included in equation~(\ref{eq:2pcf_w_cov_bin}). For two wedges, the sum can be safely truncated after $\ell = 4$ without changing the results on a significant level. In case of three wedges, the convergence at an appropriate level is reached by including terms up to $\ell = 6$.}
 \label{fig:minerva_2PCF_w_cov_maxl}
\end{figure}

We consider 2PCF clustering wedges as defined in equation~(\ref{eq:xi_w}), and denote the measurement of $\xi_{\mu}^{\mu+\Delta \mu}$ in the distance bin around $s_i$ by $\xi_\mu^i$.
Using the relation between wedges and Legendre moments given in equation~(\ref{eq:xi_w_from_ell}), we can find the covariance of the configuration space wedges:
\begin{equation}
 \Cov{\xi^i_\mu, \xi^j_{\mu'}} = \frac{1}{V_{s_i} \, V_{s_j} \, (\Delta \mu)^2} \int_{s_i - \Delta s/2}^{s_i + \Delta s/2} s^2 \dint s \int_{s_j - \Delta s/2}^{s_j + \Delta s/2} (s')^2 \dint s' \int_{\mu_1}^{\mu_2} \dint \tilde \mu \int_{\mu'}^{\mu' + \Delta \mu} \dint \tilde \mu'
 \sum_{\ell_1, \ell_2} \Cov{\xi_{\ell_1}(s), \xi_{\ell_2}(s')} \, \Lp[\ell_1](\mu) \, \Lp[\ell_2](\mu').
\end{equation}
By replacing the covariance of the multipoles, we finally arrive at
\begin{equation}
 \Cov{\xi^i_\mu, \xi^j_{\mu'}} = \sum_{\ell_1, \ell_2} \frac{\ii^{\ell_1+\ell_2}}{2 \pi^2} \Lpbar[\ell_1,\mu] \, \Lpbar[\ell_2,\mu'] \int_0^\infty k^2 \, \sigma^2_{\ell_1\ell_2}(k) \, \jlbar[\ell_1](ks_i) \, \jlbar[\ell_2](ks_j) \dint k,
\end{equation}
as already presented in equation~(\ref{eq:2pcf_w_cov_bin}), where the wedge-averaged Legendre polynomials $\Lpbar$ are given by equation~(\ref{eq:Lpbar}).
Evaluating the integral in that expression yields
\begin{equation}
 \label{eq:Lpbar_eval}
 \Lpbar = \frac{1}{\Delta \mu} \left[ \Lp[\ell+1](\mu + \Delta \mu) - \Lp[\ell-1](\mu + \Delta \mu) - \Lp[\ell+1](\mu) + \Lp[\ell-1](\mu) \right].
\end{equation}

The sum over $\ell_1$ and $\ell_2$ in equation~(\ref{eq:2pcf_w_cov_bin}) converges quickly.
As shown in Fig{.}~\ref{fig:minerva_2PCF_w_cov_maxl}, the two wedge dispersion is converged after including only those contributions up to $\ell_1, \ell_2 \le 4$;
the dispersion for three wedges needs only to also include $\ell_1, \ell_2 = 6$ in the sum in order to achieve convergence at an appropriate level.

\section{The precision matrix}
\label{app:prec_mat}

\begin{figure}
 \includegraphics[width=.32\columnwidth]{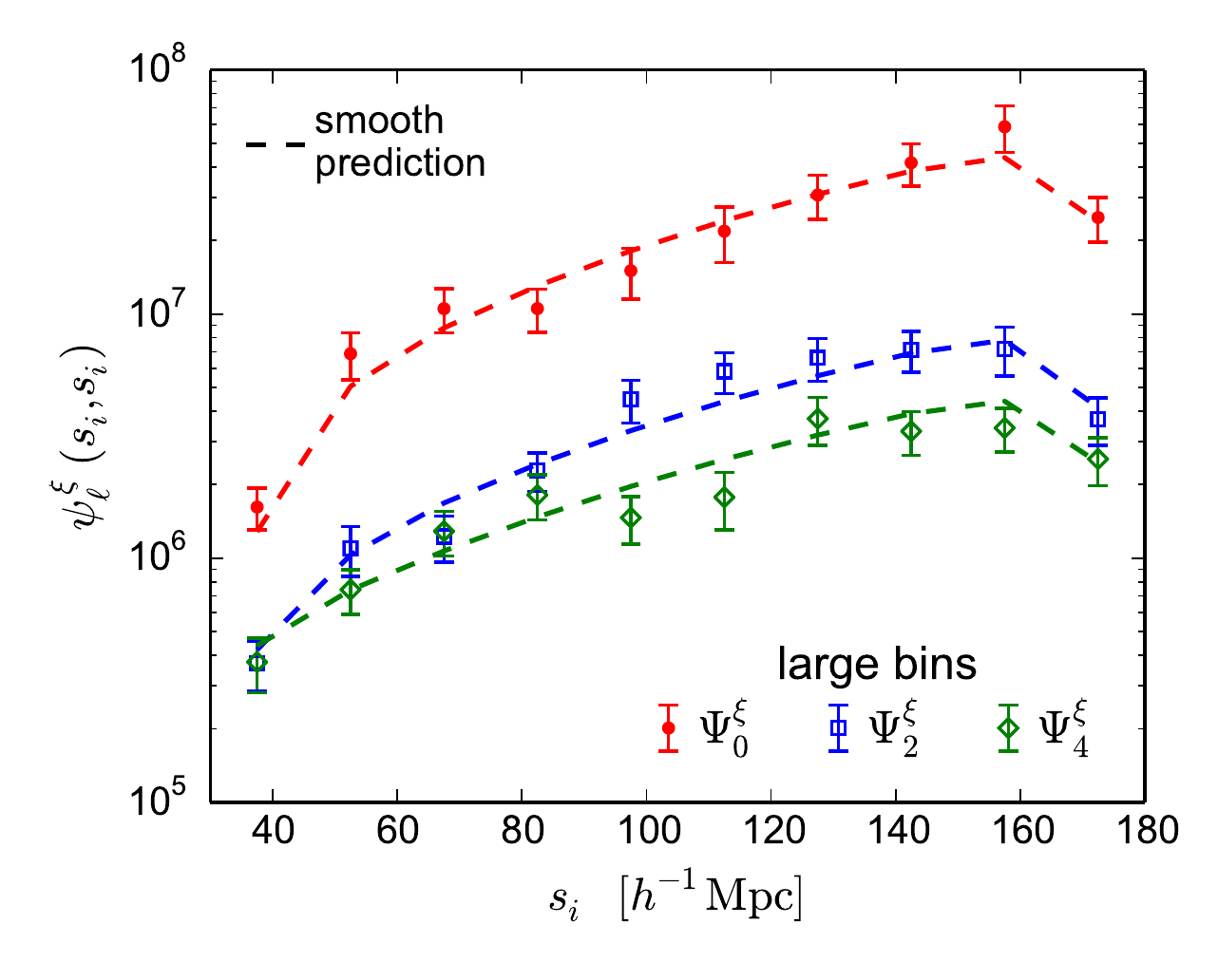}
 \includegraphics[width=.32\columnwidth]{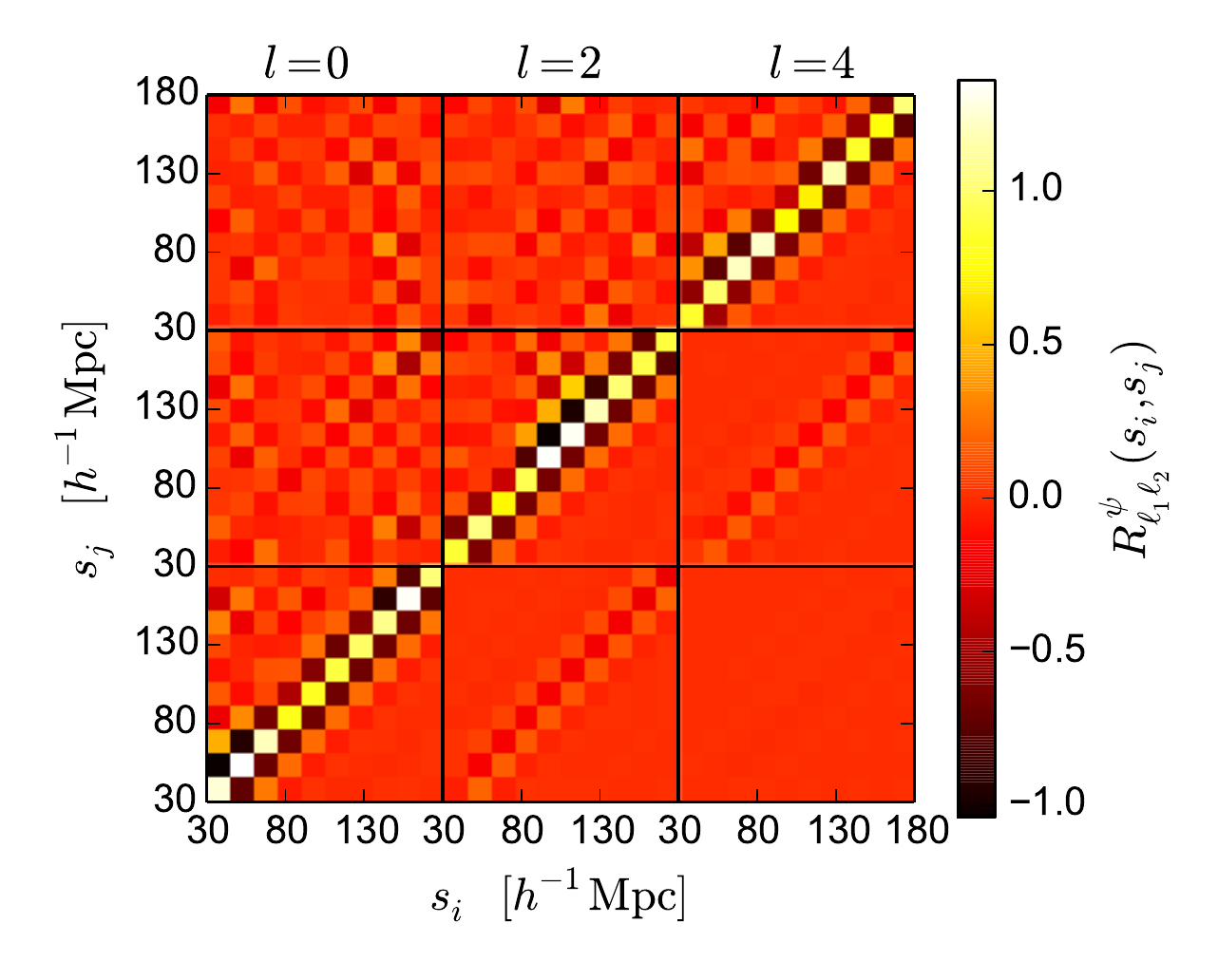}
 \includegraphics[width=.32\columnwidth]{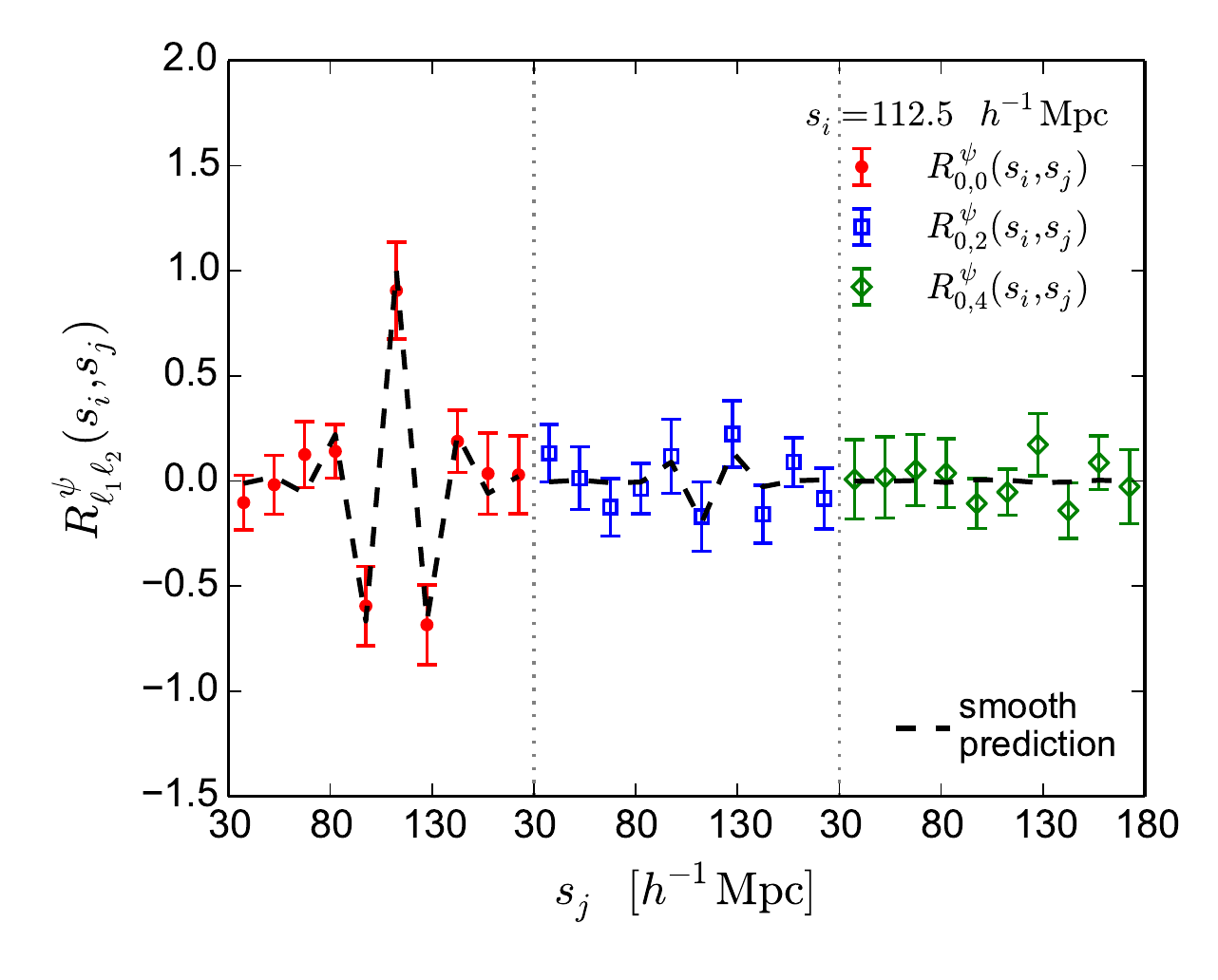}
 \caption{\emph{Left panel}: The values and error bars of the diagonal entries of the unbiased inverse data covariance matrix $\boldsymbol{\Psi}^\xi$, defined in equation~(\ref{eq:prec_mat}), of the configuration space multipole measurements where the error haven been determined with the jackknife method.
 The dashed lines show the inverse of the theoretical covariance matrix prediction with the smoothed input power spectrum.
 \emph{Center panel}: The full precision matrix of the 2PCF multipoles, $R^\psi_{\ell_1\ell_2}(k_i,k_j) = \psi^\xi_{\ell_1\ell_2}(s_i,s_j) [\psi^\xi_{\mathrm{nl},\ell_1\ell_1}(s_i,s_i) \, \psi^\xi_{\mathrm{nl},\ell_2\ell_2}(s_j,s_j))]^{-1/2}$ (normalized by the theoretical prediction), showing a complex structure and important `mixing terms'.
  \emph{Lower panel:} Cut through the precision matrix for $\ell_1 = 0$ at $s_i = 112.5 \; h^{_1} \, \unit{Mpc}$.
  The contamination from physical effects or noise not accounted for by our modelling and is within the error bars which are very large due to the inversion of a noise-contaminated covariance matrix.
  All panels show the results for $30 \; h^{-1} \, \unit{Mpc} \le s \le 180 \; h^{-1} \, \unit{Mpc}$ using $\Delta s = 10 \; h^{-1} \, \unit{Mpc}$}
 \label{fig:minerva_2PCF_l_prec}
\end{figure}

Evaluation of the Gaussian likelihood via the $\chi^2$ function given in equation~(\ref{eq:chi_sqr}) needs an estimate of the precision matrix. As pointed out in section~\ref{sec:validation_with_simulations}, the covariance matrix ${\mathbfss C}$ estimated from an ensemble of realizations is affected by noise, resulting in a biased estimate of the precision matrix \citep{Hartlap:2006kj}.
This bias can be removed by a rescaling of the inverse covariance matrix,
\begin{equation}
 \label{eq:prec_mat}
 \boldsymbol{\psi} = (1 - D) \, {\mathbfss C}^{-1}, \quad \text{where} \quad D = \frac{\Nbins + 1}{\Nmock - 1}.
\end{equation}
In order not to be noise-dominated, we require $D < 0.5$.
The only binning scheme (using three projections in $\mu$) that fulfils this requirement is the `large' scheme in configuration space ($\Delta s = 15 \; h^{-1} \, \unit{Mpc}$, and $12$ bins per wedge/multipole measurement).
For brevity, we only discuss the multipole results, $\boldsymbol{\psi}^\xi_\ell = (1 - D) \, ({\mathbfss C}^\xi_\ell)^{-1}$, where ${\mathbfss C}^\xi_\ell$ has elements $C_{\ell_1 \ell_2}(s_i, s_j)$.
Further, the axis-averaging of the measurements of the redshift-space two-point statistics is abandoned because equation~\ref{eq:prec_mat} would need to be modified in that case.

In analogy to the covariance notation, we write the elements of the precision matrix as $\psi^\xi_{\ell_1\ell_2}(s_j,s_j)$. The theoretical prediction, $\psi^\xi_{\mathrm{nl},\ell_1\ell_2}(s_j,s_j)$, is obtained from the Gaussian prediction for the multipole covariance using the smoothed input power spectrum (and by setting $D=0$ in equation~(\ref{eq:prec_mat}) due to the absence of sampling noise).
Here we focus only on the non-linear prediction for illustration.
The prediction from the linear input power spectrum is very similar, as already discussed in section~\ref{sec:validation_with_simulations}.

The measured and predicted diagonal entries of the precision matrix are shown in Fig{.}~\ref{fig:minerva_2PCF_l_prec}.
The error bars are determined with the jackknife technique (and for each jackknife estimate, the inverse covariance matrix is rescaled with a modified correction factor $1 - D$ accounting for the removed realization by $\Nmock \to \Nmock - 1$).
Due to the fact that the covariance matrix is only poorly determined, the errors on the precision matrix are very large (up to ca{.} 20\% after the application of the correction factor).
Within these errors, the model predictions $\psi^\xi_\mathrm{nl}$ do not show significant deviations.

In order to show the accuracy of the modelling for the off-diagonal terms, we plot the full precision matrix normalized by the theoretical prediction,
\begin{equation}
 R^\psi_{\ell_1\ell_2}(k_i,k_j) = \psi^\xi_{\ell_1\ell_2}(s_i,s_j) \left[ \psi^\xi_{\mathrm{nl},\ell_1\ell_1}(s_i,s_i) \, \psi^\xi_{\mathrm{nl},\ell_2\ell_2}(s_j,s_j)) \right]^{-\frac 1 2},
\end{equation}
in the center panel.
In order to better visualize the complex structure a cut through this matrix is shown in the right panel.
The sub-diagonal entries are negative and `mirror' the diagonal entries.
A similar structure is found for terms mixing the monopole and quadrupole as well as the quadrupole and hexadecapole entries.
Further away from the block diagonals, the structure is noise-dominated and deviations between data and theory are largely below the noise level.

%%%%%%%%%%%%%%%%%%%%%%%%%%%%%%%%%%%%%%%%%%%%%%%%%%

% Don't change these lines
\bsp	% typesetting comment
\label{lastpage}
\end{document}